\documentclass[twocolumn,showpacs,preprintnumbers,amsmath,amssymb,
groupedaddress,prc,floatfix]{revtex4}
\usepackage{mathrsfs}
\usepackage{longtable}
\usepackage{graphicx}
\usepackage{graphics}
\usepackage{epsfig}
\usepackage{dcolumn}
\usepackage{rotating}

\begin{document}

\title{Aspects of Single Particle Excitations and Collectivity in $^{69}$Ga}

\author{Anil Sharma}
\affiliation{UGC-DAE Consortium for Scientific Research, Kolkata Centre, Kolkata 700098, India}
\author{S. Nandi}
\affiliation{UGC-DAE Consortium for Scientific Research, Kolkata Centre, Kolkata 700098, India}
\author{S. Samanta}
\altaffiliation{Present Affiliation: Department of Physics, ADAMAS University, Kolkata 700126, India}
\affiliation{UGC-DAE Consortium for Scientific Research, Kolkata Centre, Kolkata 700098, India}
\author{S. Kundu}
\affiliation{UGC-DAE Consortium for Scientific Research, Kolkata Centre, Kolkata 700098, India}
\author{A. Das}
\affiliation{UGC-DAE Consortium for Scientific Research, Kolkata Centre, Kolkata 700098, India}
\author{ S. S. Ghugre}
\affiliation{UGC-DAE Consortium for Scientific Research, Kolkata Centre, Kolkata 700098, India}
\author{S. S. Tiwary}
\affiliation{Department of Physics, Manipal University Jaipur, Rajasthan 303007, India}
\author{I. Bala}
\affiliation{Inter University Accelerator Centre, New Delhi 110067, India}
\author{R. P. Singh}
\affiliation{Inter University Accelerator Centre, New Delhi 110067, India}
\author{S. Muralithar}
\affiliation{Inter University Accelerator Centre, New Delhi 110067, India}
\author{R. Raut}
\altaffiliation{Corresponding Author: rajarshi.raut@gmail.com}
\affiliation{UGC-DAE Consortium for Scientific Research, Kolkata Centre, Kolkata 700098, India}
\date{\today}

\begin{abstract}
The excitation scheme of the $^{69}$Ga ($Z = 31, N = 38$) nucleus has been studied following its population in $^{59}$Co($^{13}$C,2pn) reaction at $E_{lab} = 45, 50$ MeV, and using an array of 
Compton suppressed HPGe clover detectors as the detection system. The existing level scheme has been considerably extended with identification of new $\gamma$-ray transitions 
and their multipolarity assignments. The level energies have been calculated in the framework of the large basis shell model and their overlap with the experimental values 
has been satisfactory, subject to the choice of the interaction. 
The band structures identified in the nucleus have been characterized with the Moment of Inertia (MOI) and aligned angular momentum, 
and compared to those of the similar structures in the neighboring isotopes. The shapes corresponding to these bands have been probed in the 
light of their Total Routhian Surface (TRS) that exhibited varied deformation characteristics, such as prolate and $\gamma$-softness, associated with the individual sequences.
Further, evidence of strong octupole correlation has been identified from an E3 transition between bands of opposite parities. The study comprehensively brings forth multiple aspects of single particle and collective 
characteristics in the level structure of $^{69}$Ga. 
\end{abstract}


\maketitle

\section{Introduction}

Investigating the evolution of nuclear structure with increasing number of valence nucleons outside a magic core might have intriguing dividends. 
For instance, the results could illustrate the transition from single particle excitations to collective degrees of freedom 
underlying the excitation schemes, or interplay between the two paradigms, and actually facilitate a microscopic insight of the nucleon-nucleon interaction \cite{Sor02}.
Such studies are also pertinent to the developments in the shell model framework. The advancements in the computational wherewithal
through the recent decades have helped in expanded and extensive applications of the shell model calculations in nuclear structure studies.
The associated endeavors have noticeably progressed from nuclei in the immediate vicinity of the magicities \cite{Sam18,Sam19,Cha23} to even those around the mid-shell \cite{Sch09,Zho21} region and
have been directed at interpreting the level structures exhibiting signatures of single particle excitations as well as transitional characteristics 
and collective behavior \cite{Aya22,Aya23}. The sustenance of such accomplishments, however, is critically dependent on the availability of spectroscopy data
that can be used to constrain and validate the aforementioned theory. \\

The level structure of nuclei to the \textquotedblleft north-east\textquotedblright of the doubly-magic $^{56}$Ni-core ($Z, N = 28$) have been
much pursued in recent years and known to exhibit the evolving traits of excitation mechanisms with increasing number of valence nucleons \cite{Kay25} and/or increasing 
excitation energy and spin \cite{Bha23}. The Ga ($Z = 31$) isotopic chain, with three protons outside the closure at $Z = 28$, manifests
such transitional character along with interplay between the single particle and the collective aspects of nuclear excitations. The level scheme of $^{67}$Ga, for instance, 
has recently been reported \cite{Zho24} with positive and negative parity band structures and evidence of octupole correlation; the results were interpreted in the light of the reflection asymmetric triaxial 
particle rotor model as well as microscopic calculations for validating the deformation / shape of the nucleus. The excitation scheme of the $^{71}$Ga isotope \cite{Zho21}, also 
characterized with multiple band structures,
too has been proposed as resulting from coupling of the single particle excitations with the positive parity states of the neighboring even-even core ($^{70}$Zn).
In conjunction, large basis shell model calculations have been carried out \cite{Zho21} for the excited states of $^{71}$Ga, following which the levels 
were assigned particle configurations based on protons in the $p_{3/2}$ and $f_{5/2}$ orbitals and neutron occupancy in the $fpg$ model space consisting of 
the $p_{3/2}$, $f_{5/2}$, $p_{1/2}$ and $g_{9/2}$ orbitals. It emerged that the states of higher spins were built on neutron pair excitations
into the $g_{9/2}$ orbital that presumably translate into the deformation characteristics of the nucleus.
Band structures built on $g_{9/2}$, $f_{5/2}$ and $p_{3/2}$ states were systematically investigated \cite{Ste09} in the neutron-rich $^{71,73,75,77}$Ga isotopes 
and the high-spin levels were, once again, ascribed to the coupling of the odd proton to the yrast states of the respective even-even (Zn) core.
Alternatively, it is worth mentioning that the shell model calculations were earlier reported \cite{Yos08} for neutron-rich isotopes of Ga, Ge, As and Se, and the
overlap between the theoretical and the experimental level energies were largely satisfactory, at least for the states below around 1.5~MeV.   
The prospects of microscopic understanding of excitations in place, the shell model calculations have continued to be used in a number of spectroscopic 
endeavors on nuclei with considerable number ($\gtrapprox$ 10) of nucleons in the $fpg$ valence space. 
Sporadic examples of such efforts include the investigation of band structures in the $^{71}$Ge ($Z = 32$) nucleus \cite{Kay25} and 
study on the collectivity of low-lying states in $^{70,72,74}$Zn \cite{Lou13} isotopes. 
The application of shell model calculations to interpret new and updated data continues to be a pertinent exercise that is envisaged to 
facilitate the progress of the theoretical framework and the (microscopic) understanding of the level structures it entails. \\ 

This work is about experimental investigation of the excitation scheme of $^{69}$Ga ($Z = 31, N = 38$) nucleus and its interpretation 
in the light of the large basis shell model calculations, as well as analyzing the evidence(s) of collectivity therein. 
The previous \cite{Bak82,Par81,Par78,Rin74} studies on the nucleus were 
primarily based on the use of light ($\alpha, p, n$, $^7$Li) projectiles for populating states of modest excitations that were 
probed with limited number of detectors in the experimental setup. The last study on the nucleus by Bakoyeorgos {\it{et al.}} \cite{Bak82} reported its level structure 
up to an excitation energy of $\approx$ 4.5~MeV and spin $\approx$ 8$\hbar$, along with level lifetime measurements. The observed 
level structure was largely ascribed to particle-core coupling of type $g_{9/2},f_{5/2}~\otimes~J=0^+,2^+,4^+,6^+$ while considering 
substantial collectivity associated with the levels de-exciting by $E2$ transitions. The present investigation addresses the excitation scheme
of the nucleus using advanced experimental facilities, based on large and efficient array of high-resolution $\gamma$-ray detectors, along with
the shell model framework, implemented with larger model space and updated interactions, for interpreting the results.   
The outcome, in a larger context, should translate into a validation for application of microscopic models in the understanding of the level structure
of transitional nuclei characterized with considerable collectivity. 
It is also expected that the results would contribute to the perspectives on the evolving features of the level structure with
increasing number of nucleons away from the shell closures.

\section{Experimental Details and Data Analysis}

\begin{figure}
\includegraphics[angle=0,scale=.35,trim=0.0cm 0.0cm 0.0cm 0.0cm,clip=true]{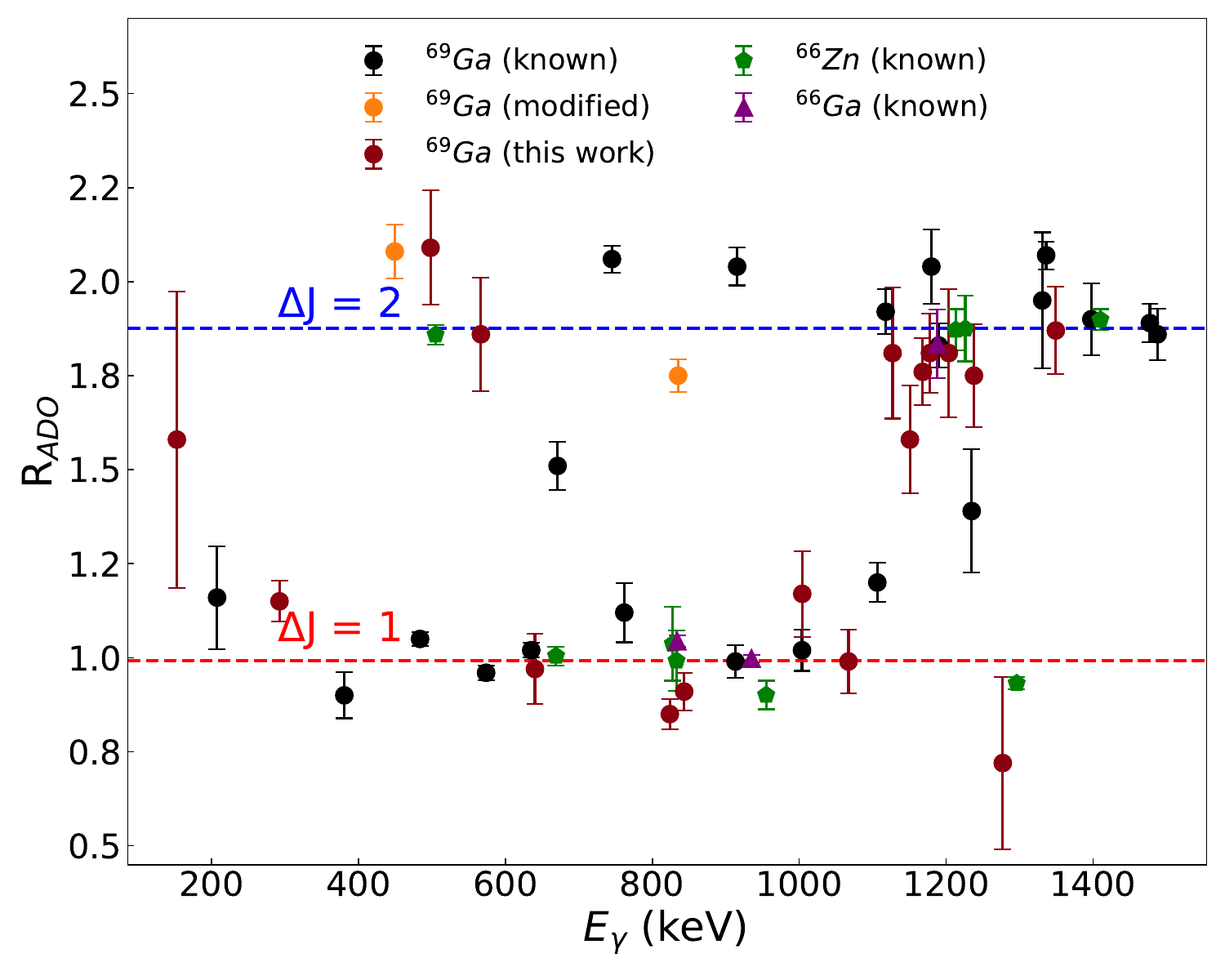}
\caption{\label{fig1}$R_{ADO}$ for transitions of $^{69}$Ga along with those of other nuclei, following the present analysis. 
The previously known pure transitions of other nuclei have been used to extract the reference values for the quadrupole and the dipole transitions.}
\end{figure}

\begin{figure}
\includegraphics[angle=0,scale=.28,trim=0.0cm 0.0cm 0.0cm 0.0cm,clip=true]{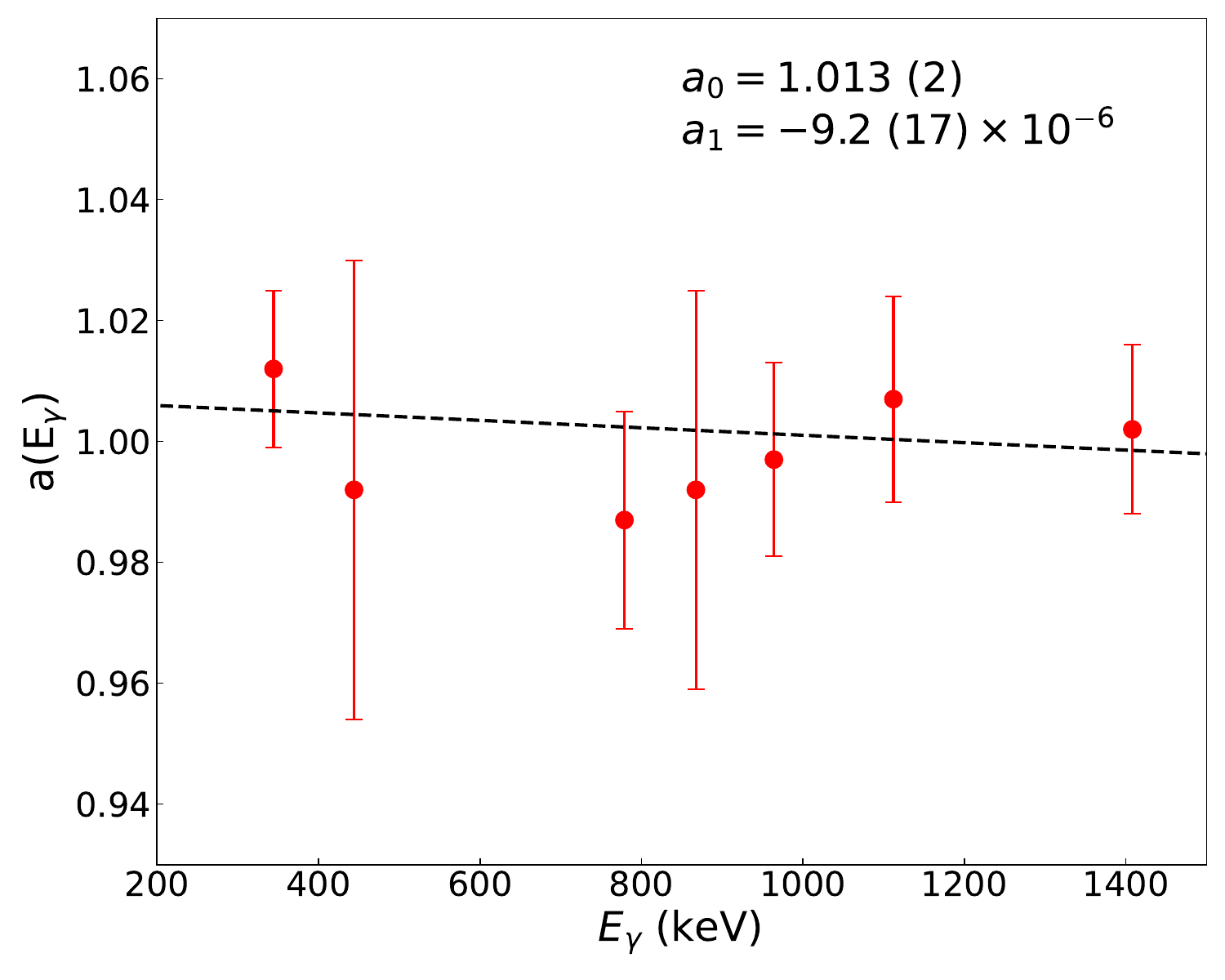}
\includegraphics[angle=0,scale=.28,trim=0.0cm 0.0cm 0.0cm 0.0cm,clip=true]{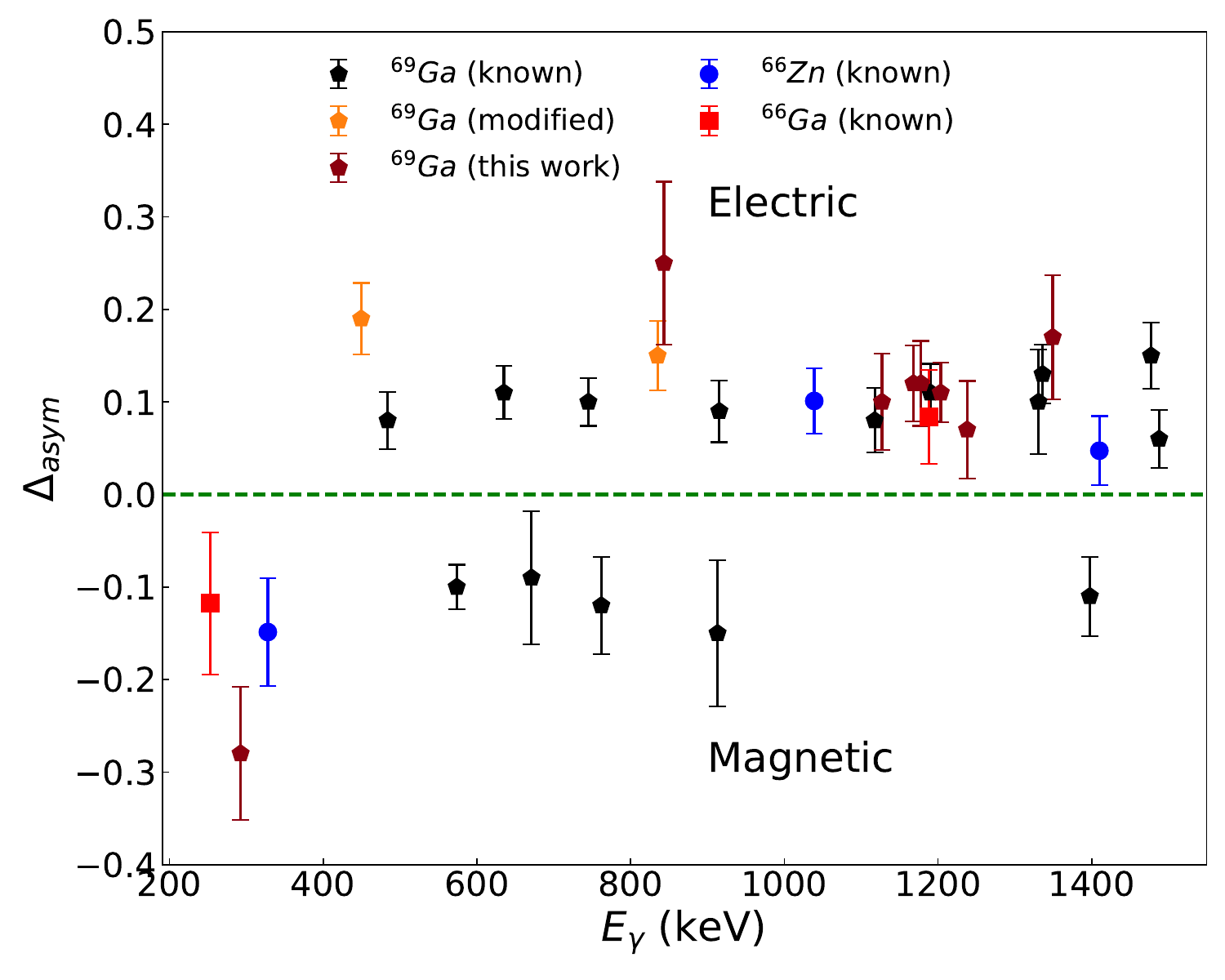}
\includegraphics[angle=0,scale=.28,trim=0.0cm 0.0cm 0.0cm 0.0cm,clip=true]{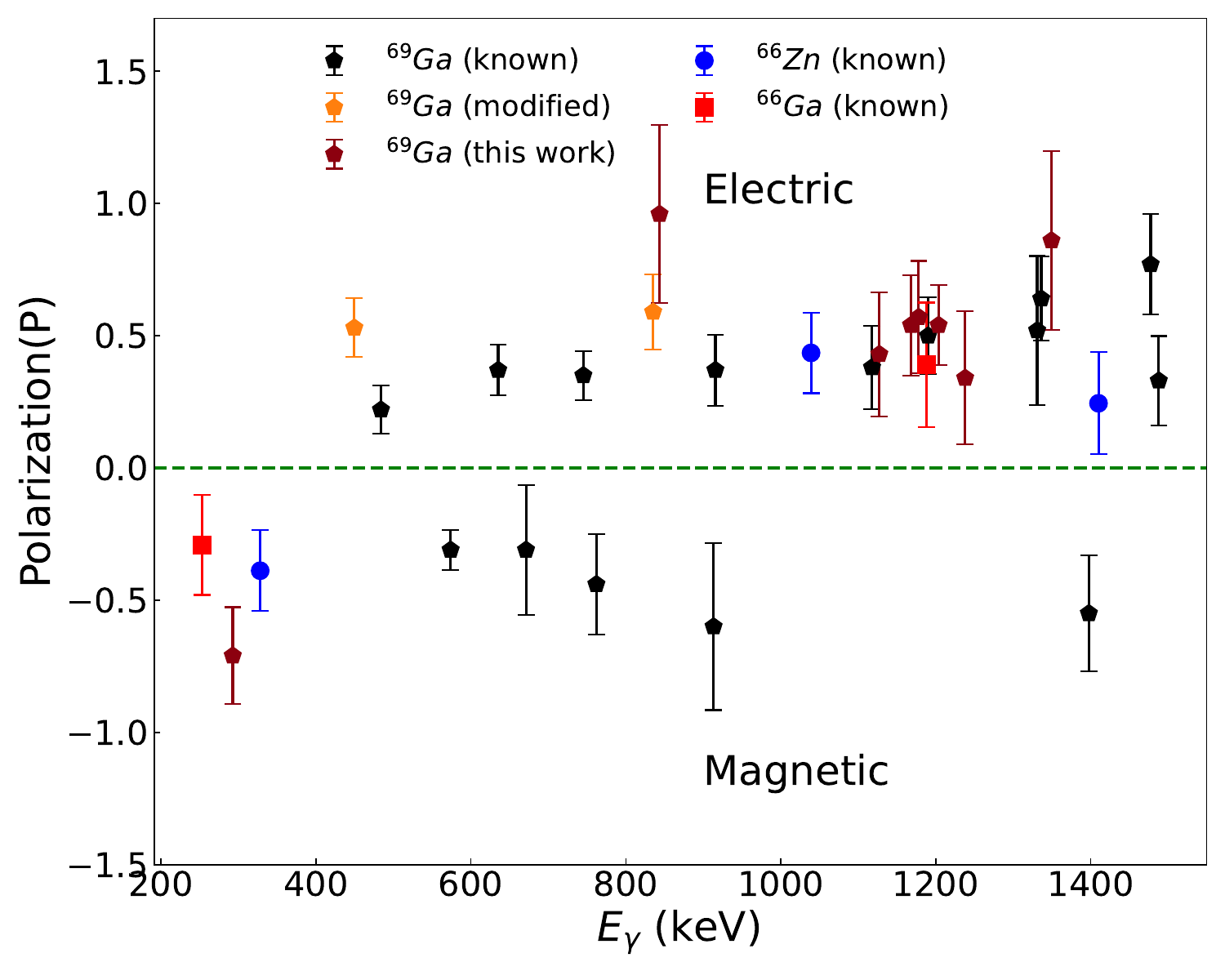}
\caption{\label{fig2}(a) Geometrical asymmetry ($a$) of the present setup for different $\gamma$-ray energies of $^{152}$Eu source.
The plot includes the linear fit of the data and the parameters ($a_0, a_1$) therefrom.
(b) Polarization asymmetry ($\Delta$) of transitions of $^{69}$Ga, following the present analysis. The plot includes $\Delta$ of 
previously known transitions of other nuclei populated in the same experiment. (c) Linear 
polarization ($P$) of transitions of $^{69}$Ga along with those of other nuclei populated in the same experiment. 
The $\Delta$ and $P$ values for selected transitions of other isotopes, that were 
produced in the same experiment, are included for validation.}
\end{figure}

\begin{figure*}
\includegraphics[angle=0,scale=0.75,trim=1.0cm 4.0cm 3.0cm 4.0cm,clip=true]{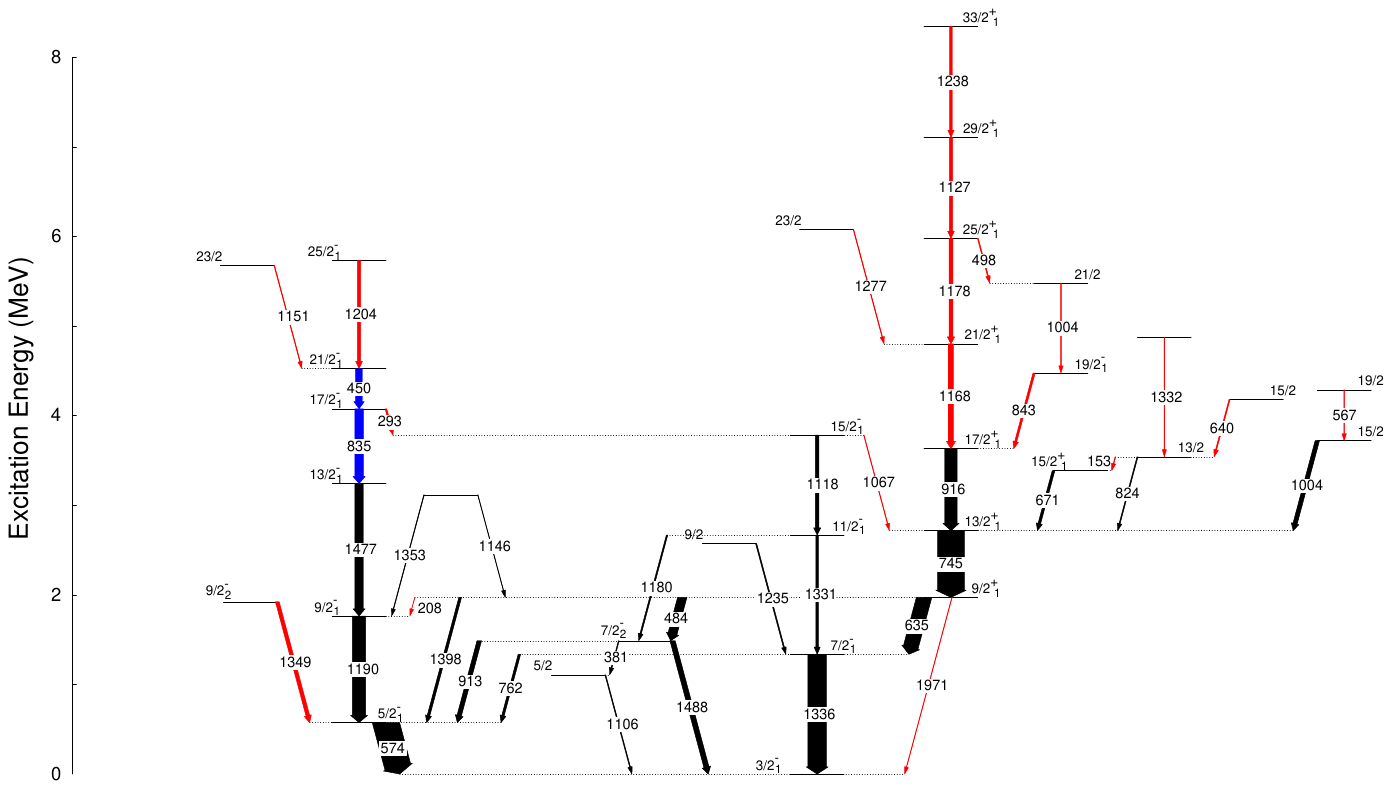}
\caption{\label{fig3} Excitation scheme of $^{69}$Ga, following the present investigation. The $\gamma$-ray 
transitions that have been newly identified in this work are marked in red. 
The transitions marked in blue are those for which the spin-parity of the initial level has been modified in this work.}
\end{figure*}

\begin{figure}
\includegraphics[angle=0,scale=.30,trim=0.0cm 0.0cm 0.0cm 0.0cm,clip=true]{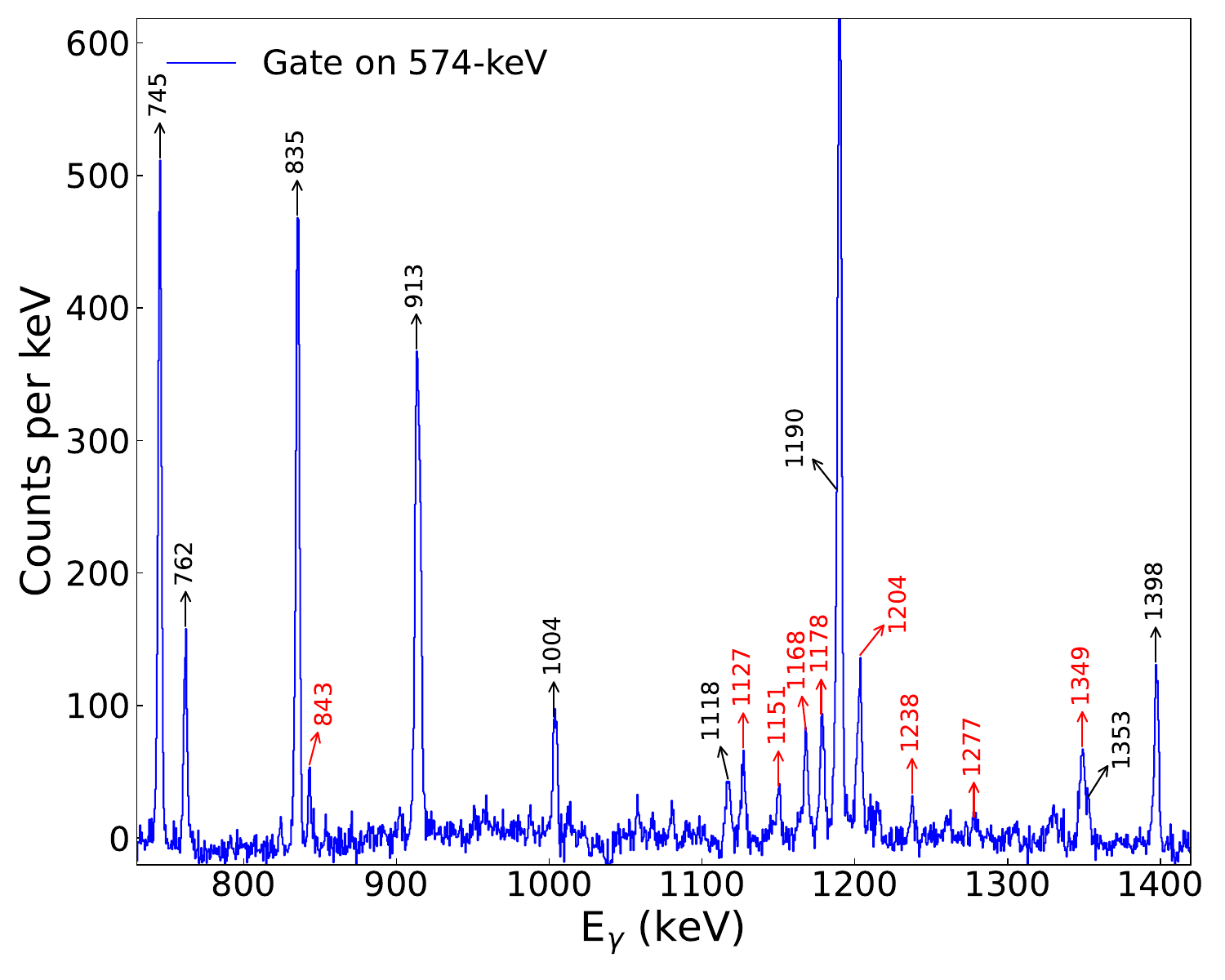}
\includegraphics[angle=0,scale=.30,trim=0.0cm 0.0cm 0.0cm 0.0cm,clip=true]{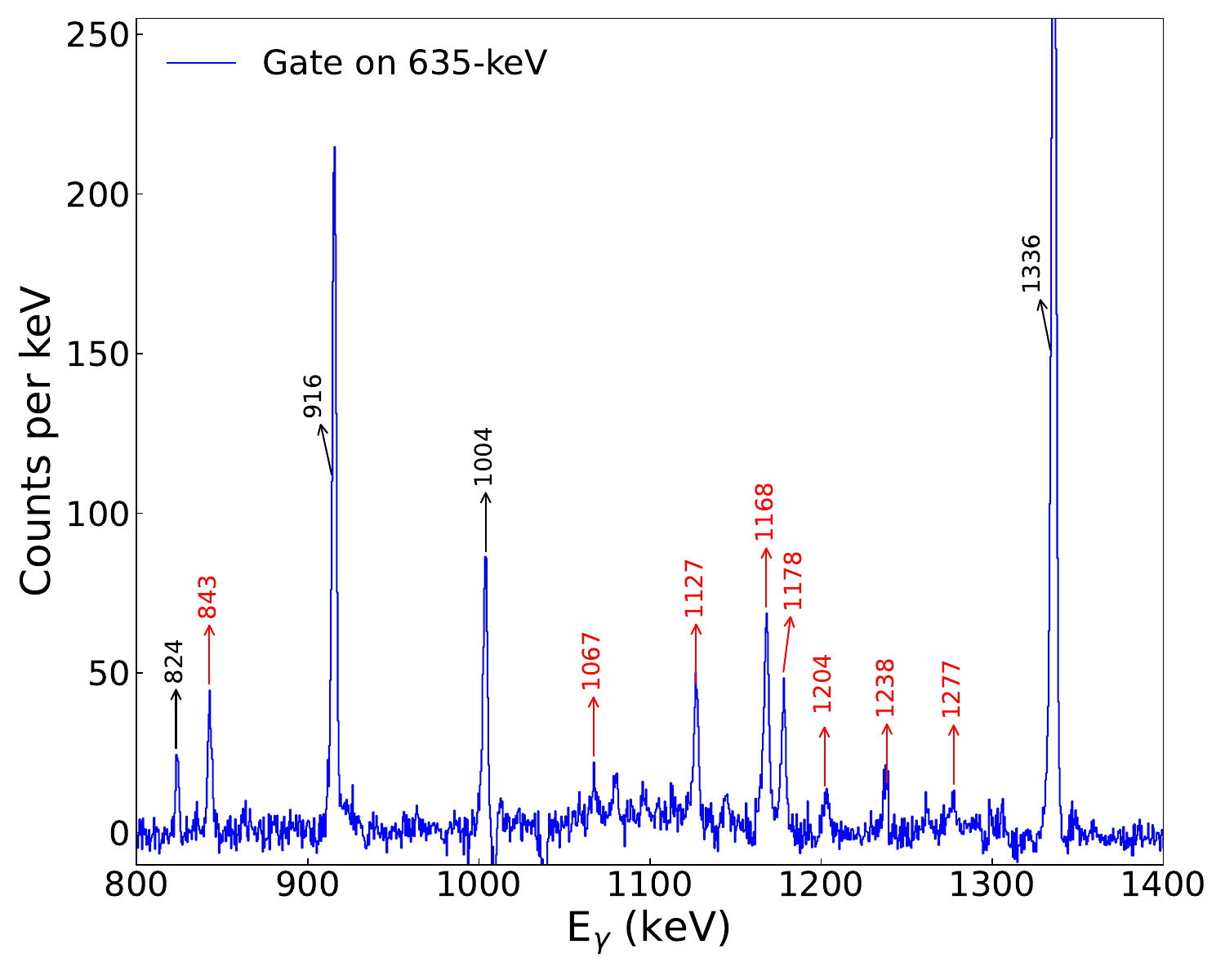}
\includegraphics[angle=0,scale=.30,trim=0.0cm 0.0cm 0.0cm 0.0cm,clip=true]{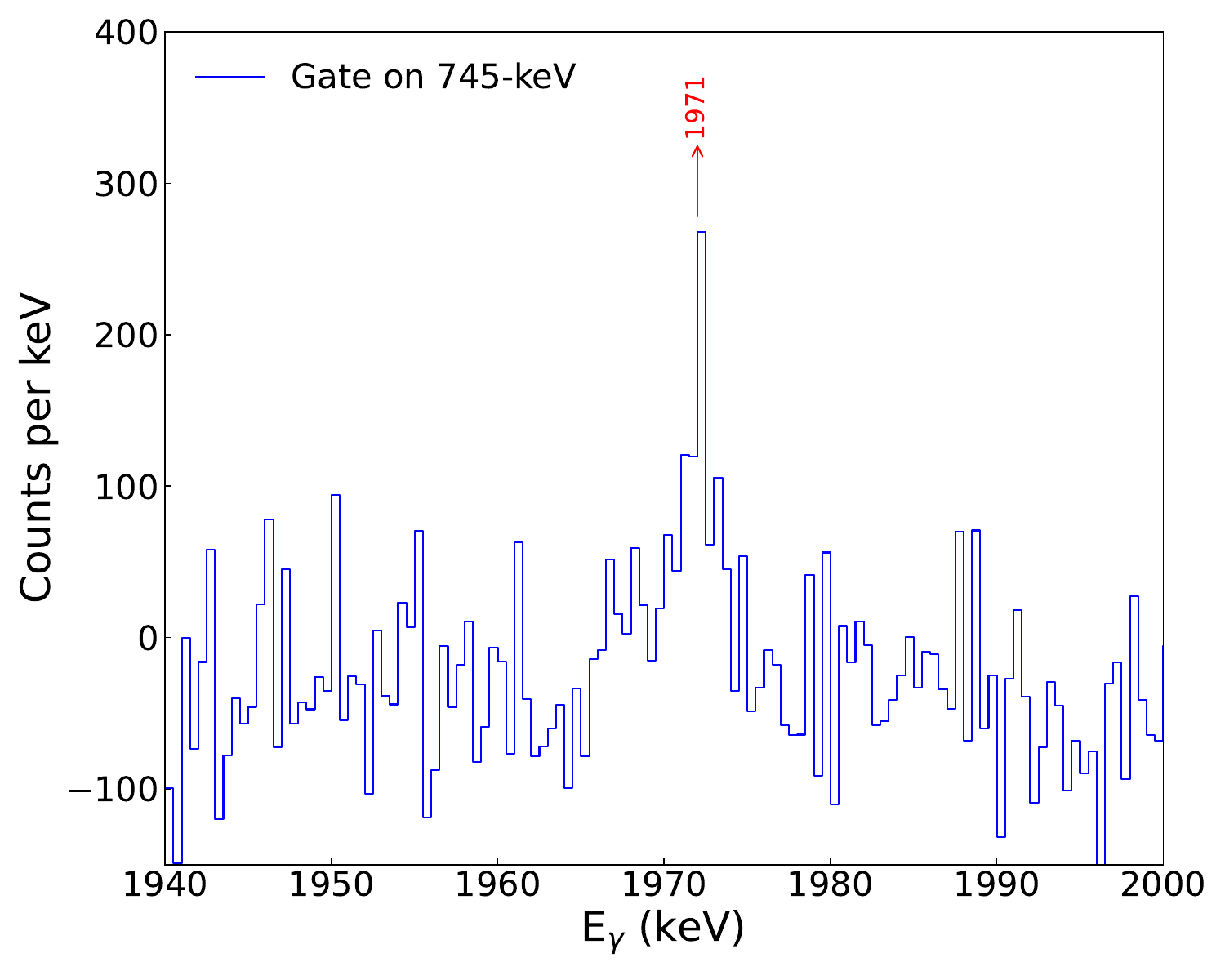}
\caption{\label{fig4}Representative spectra corresponding to gates applied on transitions of $^{69}$Ga, projected out of $\gamma$-$\gamma$ matrix. The gating transitions are indicated in the inset of the respective spectrum. The new transitions identified in the respective gate are marked in red. The gate on the 745-keV transition is directed at illustrating the 1971-keV transition that has been (tentatively) identified as an E3 one and facilitates the proposition of octupole correlation in the nucleus.}
\end{figure}

\begin{figure}
\includegraphics[angle=0,scale=.30,trim=0.0cm 0.0cm 0.0cm 0.0cm,clip=true]{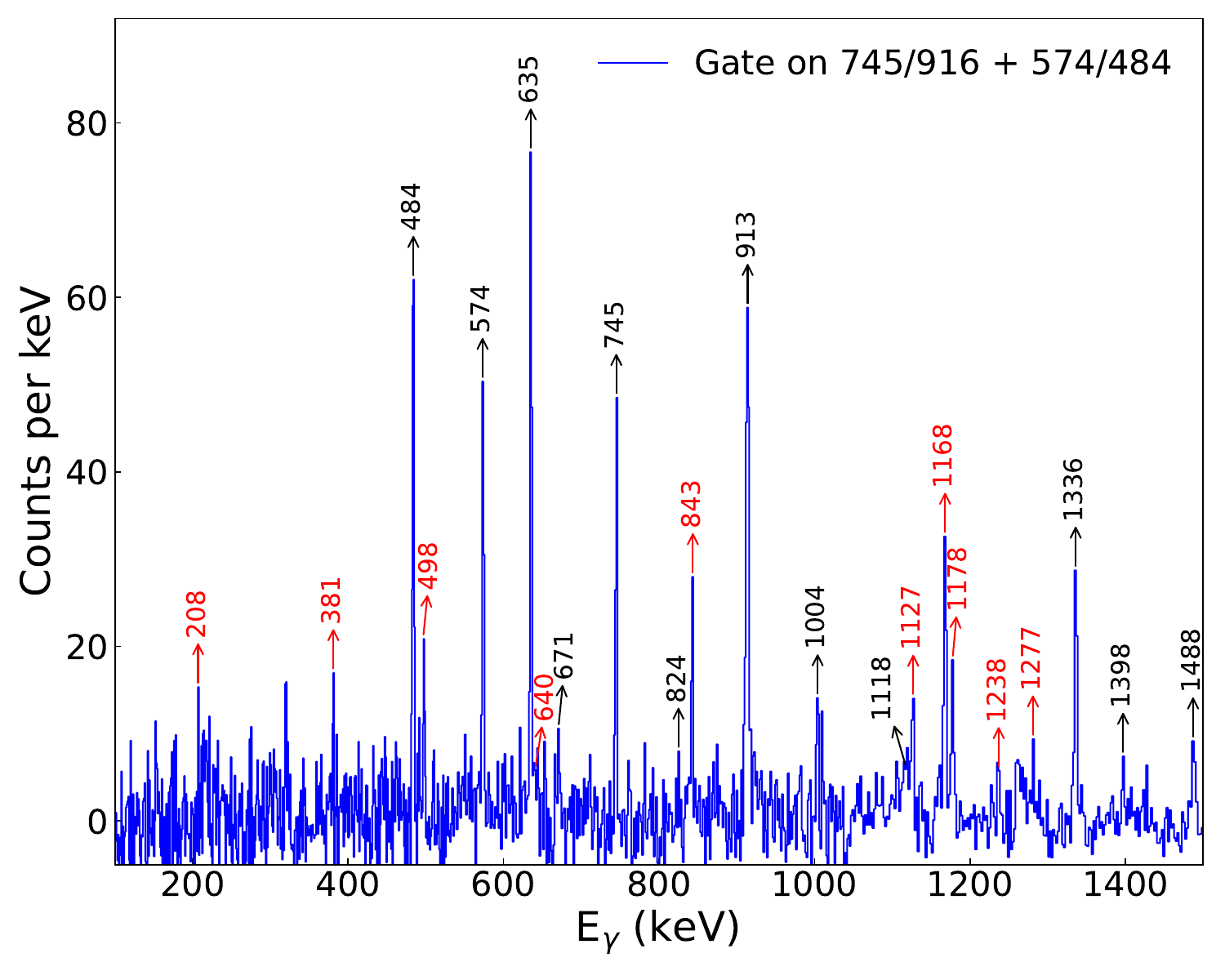}
\caption{\label{fig5}Representative spectrum projected out of $\gamma$-$\gamma$-$\gamma$ cube corresponding to double gates set on transitions of $^{69}$Ga. The same facilitates more stringent validation of the $\gamma$-$\gamma$ coincidences identified for the nucleus. The new transitions are indicated in red.}
\end{figure}

\begin{figure}
\includegraphics[angle=0,scale=.30,trim=0.0cm 0.0cm 0.0cm 0.0cm,clip=true]{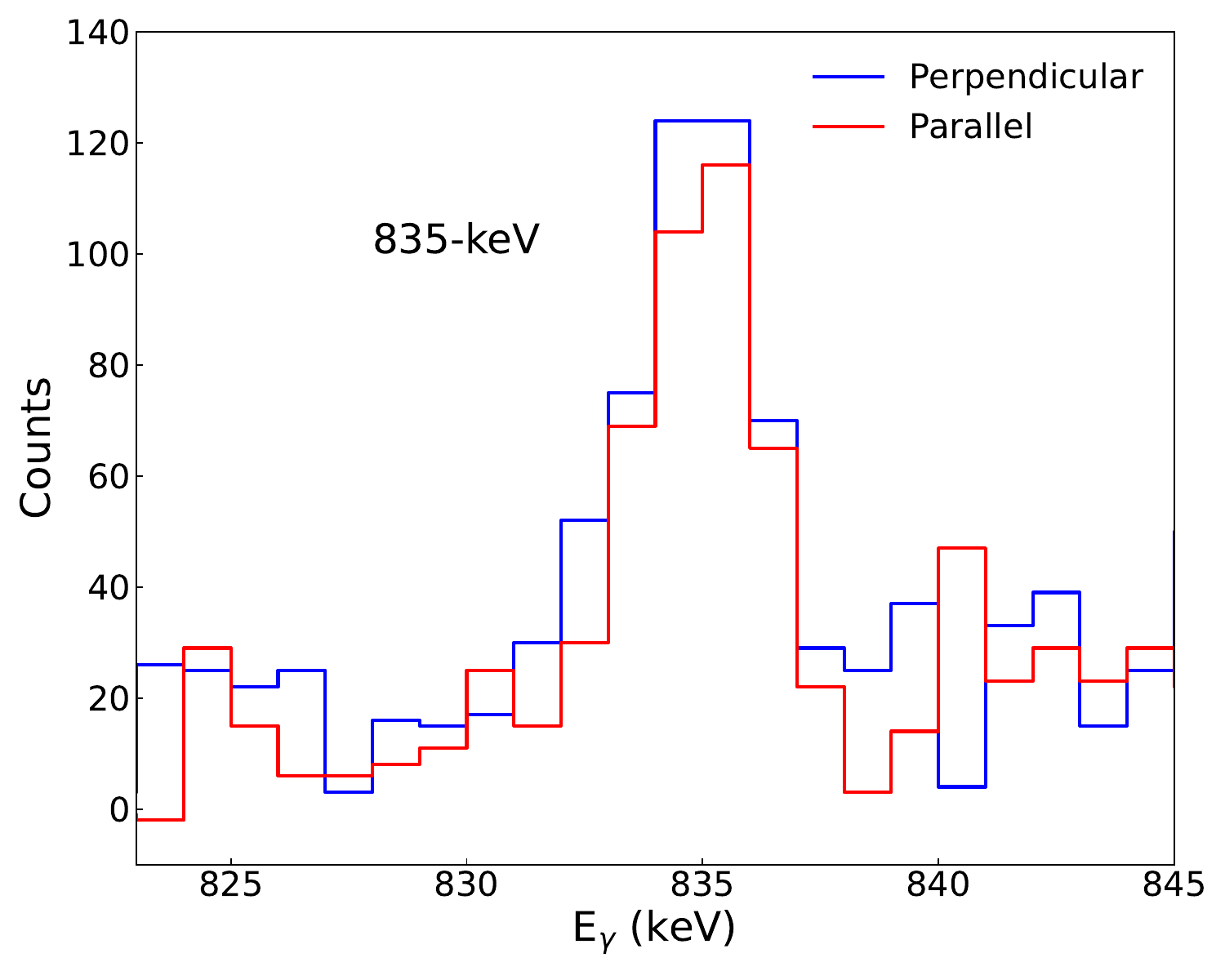}
\caption{\label{fig6}Spectra of 835-keV transition peak corresponding to the perpendicular and the parallel scattering events in the HPGe clover detectors at 90$^o$.}
\end{figure}

The excited states of $^{69}$Ga nucleus were populated following $^{59}$Co($^{13}$C,$2pn$) reaction at $E_{lab}$ = 45, 50 MeV.
The target was 5.2~mg/cm$^2$ thick mono-isotopic $^{59}$Co evaporated on a 4~mg/cm$^2$ thick tantalum (Ta) backing. 
The $^{13}$C beam was delivered by the 15 UD Pelletron at the Inter University Accelerator Centre (IUAC), New Delhi.
The Indian National Gamma Array (INGA) \cite{Mur10} consisting of 12 Compton suppressed HPGe clover detectors, during the experiment,
was used as the detector system. The detectors were positioned at 32$^\circ$ (3), 57$^\circ$ (1), 90$^\circ$ (2), 123$^\circ$ (2), and 148$^\circ$ (4). 
Listmode data was acquired on a CAMAC based multi-crate data acquisition system running on the CANDLE software developed at IUAC \cite{Kum01}.
The event trigger was set on the coincidence multiplicity of at least two ($M_\gamma \ge 2$) Compton suppressed clover detectors.
The number of two- and higher-fold events acquired in the experiment was $\approx$ 2$\times$10$^9$. The detector array was calibrated for energy and efficiency 
using standard radioactive sources such as $^{152}$Eu and $^{133}$Ba. \\

The listmode data corresponding to $\gamma$-$\gamma$ coincidences was sorted into symmetric and asymmetric (angle dependent) matrices using the SPRINGZ \cite{Das17_2}
code developed at the UGC-DAE CSR, Kolkata Centre. The coincidence data was sorted into a $\gamma$-$\gamma$-$\gamma$ cube using SPRINGZ and the relevant subroutines 
of the INGASORT \cite{Bho01} code. The sorted data was analyzed using the RADWARE \cite{Rad95} and the CUBIX \cite{Dud25} packages. \\

As is typical of heavy-ion induced fusion-evaporation reactions, above the barrier, a number of nuclei were populated in the experiment.
The list includes isotopes of Zn ($Z = 30$), Ga ($Z = 31$) and Ge ($Z = 32$). The $\gamma$-ray transitions in level scheme of $^{69}$Ga 
were placed following the validation of $\gamma$-$\gamma$ and $\gamma$-$\gamma$-$\gamma$ coincidence relationships together with 
their intensity considerations. \\

The multipolarities of the $\gamma$-ray transitions were assigned following the determination of their $R_{ADO}$ (Ratio of Angular Distribution from Oriented nuclei) values
using,

\begin{equation}
R_{ADO} = \frac{I_{\gamma1} \ at \ 148^\circ \ (Gated \ by \ \gamma_2 \ at \ all \ angles)}{I_{\gamma1} \ at \ 90^\circ \ (Gated \ by \ \gamma_2 \ at \ all \ angles)}
\end{equation}

\noindent{The $I$ denotes the intensity of the transition $\gamma_1$, at the respective angle, as extracted from the spectrum corresponding to 
a gate on transition $\gamma_2$, detected at any angle. The angle dependent matrices used for the purpose had $\gamma$-rays detected at any angle 
on the X-axis and those detected, in coincidence, at 148$^\circ$ (90$^\circ$) on the Y-axis. In the present analysis, the $R_{ADO}$ 
value for the stretched dipole ($\Delta$J = 1) transitions is expected to be 0.99$\pm$0.01 while for the 
stretched quadrupole ($\Delta$J = 2) ones, 
it is 1.88$\pm$0.01. The reference numbers were derived from the weighted average of the $R_{ADO}$ values of previously known transitions 
of other nuclei ($^{66}$Zn, $^{66}$Ga) produced in the same experiment. 
The particular transitions used for the purpose include the dipole \cite{nndc}transitions 669-keV, 5$^-$$\rightarrow$4$^+$; 828-keV, 10$^+$$\rightarrow$9$^-$; 833-keV, 2$^+$$\rightarrow$2$^+$; 955-keV, 8$^+$$\rightarrow$7$^-$ and 
1296-keV, 5$^-$$\rightarrow$4$^+$ of $^{66}$Zn, and 834-keV, 5$^+$$\rightarrow$4$^+$; 935-keV, 5$^+$$\rightarrow$4$^+$ of $^{66}$Ga. 
The reference value for stretched quadrupole \cite{nndc} transitions follow the $R_{ADO}$ of 505-keV, 7$^-$$\rightarrow$5$^-$; 1213-keV, 9$^-$$\rightarrow$7$^-$; 1226-keV, 12$^+$$\rightarrow$10$^+$; 1410-keV, 4$^+$$\rightarrow$2$^+$
transitions of $^{66}$Zn and 1188-keV, 9$^+$$\rightarrow$7$^+$of $^{66}$Ga. 
Fig.~1 is the plot of the $R_{ADO}$ values determined for different $\gamma$-ray transitions, observed in this study.} \\

The electric/ magnetic nature of the $\gamma$-ray transitions could be assigned following the measurement of their linear polarization while using  
the geometrical merit of the HPGe clover detector with four crystals closely mounted on a common cryostat. The arrangement is known \cite{Duc99} 
to facilitate measurement of polarization asymmetry with each crystal operating as a scatterer and the ones adjacent to it, in the perpendicular and parallel 
directions, as absorbers. The polarization asymmetry, for a given $\gamma$-ray transition, is defined by,

\begin{equation}
\Delta = \frac{aN_\perp \ - \ N_\parallel}{aN_\perp \ + \ N_\parallel}
\end{equation}

\noindent{where $N_\perp$ ($N_\parallel$) are respectively the number of photons of the $\gamma$-ray
that are scattered perpendicular (parallel) to the reference plane. The reference plane is identified as one containing 
the beam direction and the direction of emission of the $\gamma$-ray. 
The electric (magnetic) transitions are characterized by positive (negative) values of $\Delta$ while its value is 
near 0 for the mixed ones. 
The $N_\perp$ ($N_\parallel$) for $\gamma$-rays is extracted from a
matrix constructed with the perpendicular (parallel) scattering events in the detectors 
at 90$^o$ on the Y-axis and the coincident $\gamma$-rays detected at any angle on the X-axis. 
The coincidences simply ensure identification of the $\gamma$-ray transitions of interest. 
The $a$ in the aforesaid equation is the geometrical asymmetry inherent to the detection setup. 
It is determined from the difference between $N_\perp$ and $N_\parallel$ 
for the $\gamma$-rays of unpolarized radioactive sources, such as $^{152}$Eu, and using $a = N_\parallel/N_\perp$.
The typical plot of $a$ for different $\gamma$-ray energies, as observed in the present setup, is 
illustrated in Fig.~2(a). The data for $a$ is conventionally fitted with a first-order polynomial ($a_0~+~a_1x$) 
wherefrom $a$ is assigned the value of the dominant offset ($a_0$) term; $a_1$ is typically negligible (10$^{-6}$ in this case). In the present analysis, $a$ = 1.013$\pm$0.002. 
The polarization asymmetry ($\Delta$, as defined by Eq. (2)) values of the transitions observed in this study are plotted in Fig.~2(b).
The $\Delta$ values for select transitions of nuclei other than $^{69}$Ga, and produced in the same experiment, are included as validation of the present analysis. 
These transitions, used as reference, were the same as those (aforementioned) used for determining the reference values of the $R_{ADO}$.   
The value of $\Delta$ for a transition is dependent on its energy that impacts the probability of scattering, and thus the difference between $N_\perp$ and $N_\parallel$.
The dependence is compensated by translating the polarization asymmetry ($\Delta$) into linear polarization $P$ that is defined as,

\begin{equation}
P = \frac{\Delta}{Q}
\end{equation}

\noindent{where $Q$ is the energy dependent polarization sensitivity given by,}

\begin{equation}
Q(E_\gamma) = Q_0(E_\gamma)(CE_\gamma \ + \ D)
\end{equation}

\noindent{with,}

\begin{equation}
Q_0(E_\gamma) = \frac{\alpha + 1}{\alpha^2 + \alpha + 1}
\end{equation}

\noindent{$\alpha$ being $E_\gamma/m_ec^2$, $m_ec^2$ is the electron rest mass energy. The $C$ and $D$ 
parameters used are those reported by Palit {\it{et al.}} \cite{Pal00}; $C = 0.000099 \ keV^{-1}$ and $D = 0.446$. 
Similar to the $\Delta$, the electric (magnetic) transitions are characterized by positive (negative) values of $P$ while for the mixed ones, these are near zero.
Fig. 2(c) illustrates the $P$ values for different transitions, of $^{69}$Ga and other nuclei, observed in this study. Those of other nuclei are included as validation 
of the analysis.} \\

The experimental and data analysis methodologies that are regularly practiced in nuclear structure investigations, 
and described here, led to the identification and/or validation of the excitation scheme of $^{69}$Ga nucleus. 
The results of the exercise are elaborated in the next section.

\section{Results}

The level scheme of the $^{69}$Ga nucleus, following the present work, is illustrated in Fig.~3. 
Table~1 lists the experimental energy levels and the de-exciting $\gamma$-ray transitions, along with their intensities, multipolarity assignments and the associated $R_{ADO}$ and polarization values. 
Nineteen (19) new $\gamma$-ray transitions have been identified for the nucleus, herein, and the excitation scheme has been extended upto E$_x$ $\approx$ 8.3 MeV and spin $\approx$ 16$\hbar$.
The level scheme is based on $\gamma$-$\gamma$ coincidence relationships as extracted from $\gamma$-$\gamma$ matrix and $\gamma$-$\gamma$-$\gamma$ cube, and 
$\gamma$-ray intensities. Figs.~4, 5 illustrate some of the relevant coincidences in the representative $\gamma$-ray spectra corresponding to gates on transitions of $^{69}$Ga. \\

The present results also confirm most of those reported by Bakoyeorgos {\it{et al.}} \cite{Bak82} except changes in the spin-parities of the levels 4077- and 4526-keV, respectively 
de-excited by the transitions 835- and 450-keV. The 4077-keV state was previously assigned spin-parity 15/2$^-$ that has now been revised to 17/2$^-$.  
The 4526-keV level, de-excited by the 450-keV $\gamma$-ray, was earlier assigned a tentative spin-parity 17/2$^-$ or 19/2$^-$ but has now been identified as the yrast 21/2$^-$.
These changes follow the measurement of multipolarity and the electric/magnetic nature of the transitions, carried out in the present investigation.  
For example, Fig.~6 illustrates the spectra of 835-keV transition corresponding to perpendicular and parallel scattering in the detectors at 90$^\circ$. 
The asymmetry between the two (scattering directions) translates (eq.~2) into the polarization, and thus help in assigning the electric/magnetic character, of the transition.
The spectra indicate more counts for the perpendicular scattering that results into positive (electric) value for the polarization asymmetry and, along with the 
$R_{ADO}$ value ($\Delta{J} = 2$) of the transition, justifies the E2 assignment for its multipolarity. Similar reasoning applies to the changed multipolarity (E2) of the
450-keV transition and the corresponding change in the spin-parity of the de-excited level. \\

One of the significant experimental results of the current study is the extension of the sequence based on the yrast 9/2$^+$ state.
Bakoyeorgos {\it{et al.}} \cite{Bak82} had reported the 13/2$^+$ and the 17/2$^+$ states of the sequence while confirming the 
E2 multipolarity of the de-exciting transitions following the measurement of their angular distributions. The present study has extended the 
band to 33/2$^+$ state with newly identified intraband transitions (levels) 1168- (E$_x$ = 4800-), 1178 (E$_x$ = 5978-), 1127- (E$_x$ = 7105-), and 1238- (E$_x$ = 8343-) keV.  
The transitions have all been identified as E2, following their $R_{ADO}$ and the polarization values as extracted (Table~1) in the present analysis, and the spin-parity of the de-exciting 
levels have been assigned accordingly. \\

\LTcapwidth=\textwidth
\begin{longtable*}{ccccccccccc}
\caption{\label{tab1}Details of the levels and the $\gamma$-ray transitions of $^{69}$Ga nucleus observed in the present work. The energy of a $\gamma$-ray transition is the weighted average of its value in multiple gates. The level energies ($E_i$ and $E_f$) are those following the least-square fit using the GTOL code under the Nuclear Data Services framework of the IAEA \cite{iaea}. The $D$ and $Q$ in the column for multipolarity assignments respectively represents dipole and quadrupole transitions for which the polarization measurement could not be carried out.} \\
\hline
$E_i (keV) $   & $E_{\gamma} (keV) $     & $E_f (keV) $   &$I_{\gamma}$ & $J_i^{\pi}$     & $J_f^{\pi}$&$R_{ADO}$  &$\Delta_{pol}$ & P & Multipolarity\\
\hline
\hline
\endfirsthead

\multicolumn{11}{c}%
{{ \tablename \thetable{} -- continued from previous page}} \\
\hline
$E_i (keV) $   & $E_{\gamma} (keV) $    & $E_f (keV) $   & $I_{\gamma}$ &  $J_i^{\pi}$ & $J_f^{\pi}$&$R_{ADO}$ &$\Delta_{pol}$ & P & Multipolarity\\
\hline
\endhead

\hline
\multicolumn{11}{c}{Continued in next page}\\
\hline
\endfoot
\endlastfoot

 574.0(1)& 573.8(1)  & 0.0    & 982(7)  & 5/2$^{-}$   & 3/2$^{-}$    & 0.96(2)  & -0.10(2) & -0.31(8)  & M1 \\
1106.4(1)& 1106.3(1) & 0.0    & 32(1)   & 5/2         & 3/2$^{-}$    & 1.20(5)  &          &           & D+Q \\
1336.1(1)& 761.9(1)  & 574.0  & 89(1)   & 7/2$^{-}$   & 5/2$^{-}$    & 1.12(8)  & -0.12(5) & -0.44(19) & M1+E2 \\
         & 1336.1(1) & 0.0    & 687(2)  & 7/2$^{-}$   & 3/2$^{-}$    & 2.07(4)  & 0.13(3)  &  0.64(16) & E2 \\
1487.4(1)& 380.8(2)  & 1106.4 & 32(1)   & 7/2$^{-}$   & 5/2          & 0.90(6)  &          &           & D  \\
         & 913.2(1)  & 574.0  & 164(5)  & 7/2$^{-}$   &  5/2$^{-}$   & 0.99(4)  & -0.15(8) & -0.60(32) & M1 \\
         & 1487.8(1) & 0.0    & 190(2)  & 7/2$^{-}$   & 3/2$^{-}$    & 1.86(7)  & 0.06(3)  &  0.33(17) & E2 \\
1764.0(1)& 1190.1(1) & 574.0  & 494(5)  & 9/2$^{-}$   & 5/2$^{-}$    & 1.83(6)  & 0.11(3)  &  0.50(15) & E2 \\
1923.3(1)& 1349.3(1) & 574.0  & 137(2)  & 9/2$^{-}$   & 5/2$^{-}$    & 1.87(12) & 0.17(7)  &  0.86(34) & E2 \\
1971.5(1)& 207.6(1)  & 1764.0 & 10(1)   & 9/2$^{+}$   & 9/2$^{-}$    & 1.16(14) &          &           & D+Q \\
         & 483.9(1)  & 1487.4 & 336(1)  & 9/2$^{+}$   & 7/2$^{-}$    & 1.05(2)  & 0.08(3)  &  0.22(9)  & E1 \\
         & 635.2(1)  & 1336.1 & 556(1)  & 9/2$^{+}$   & 7/2$^{-}$    & 1.02(2)  & 0.11(3)  &  0.37(10) & E1 \\
         & 1397.6(1) & 574.0  &  98(1)  & 9/2$^{+}$   & 5/2$^{-}$    & 1.90(10) & -0.11(4) & -0.55(22) & M2 \\
         & 1971.3(1) & 0.0    & 16(1)   & 9/2$^{+}$   & 3/2$^{-}$    &          &          &           & (E3) \\       
2570.9(4)& 1234.8(4) & 1336.1 & 41(1)   & 9/2         & 7/2$^{-}$    & 1.39(16) &          &           & D+Q \\
2666.9(1)& 1179.9(1) & 1487.4 & 51(1)   & 11/2$^{-}$  & 7/2$^{-}$    & 2.04(10) &          &           & Q \\
         & 1330.8(1) & 1336.1 & 91(1)   & 11/2$^{-}$  & 7/2$^{-}$    & 1.95(18) & 0.10(6)  & 0.52(28)  & E2 \\
2716.9(1)& 745.1(1)  & 1971.5 & 1000    & 13/2$^{+}$  & 9/2$^{+}$    & 2.06(4)  & 0.10(3)  & 0.35(9)   & E2 \\
3117.3(1)& 1145.6(3) & 1971.5 & 16(1)   &             & 9/2$^{+}$    &          &          &           &  \\
         & 1353.3(1) & 1764.0 & 22(1)   &             & 9/2$^{-}$    &          &          &           &  \\
3241.4(1)& 1477.3(1) & 1764.0 & 313(4)  & 13/2$^{-}$  & 9/2$^{-}$    & 1.89(5)  & 0.15(4)  & 0.77(19)  & E2 \\
3387.9(1)& 671.1(1)  & 2716.9 & 96(1)   & 15/2$^{+}$  & 13/2$^{+}$   & 1.51(6)  & -0.09(7) & -0.31(25) & M1+E2 \\
3540.8(1)& 152.9(1)  & 3387.9 & 20(1)   & 13/2        & 15/2$^{+}$   & 1.58(39) &          &           & D+Q \\
         & 823.9(1)  & 2716.9 & 37(1)   & 13/2        & 13/2$^{+}$   & 0.85(4)  &          &           & D \\
3632.4(1)& 915.5(1)  & 2716.9 & 464(2)  & 17/2$^{+}$  & 13/2$^{+}$   & 2.04(5)  & 0.09(3)  & 0.37(13)  & E2 \\
3720.7(1)& 1003.8(1) & 2716.9 & 158(1)  & 15/2        & 13/2$^{+}$   & 1.02(5)  &          &           & D \\
3784.3(1)& 1067.1(1) & 2716.9 & 19(1)   & 15/2$^{-}$  & 13/2$^{+}$   & 0.99(8)  &          &           & D \\
         & 1117.7(1) & 2666.9 & 133(2)  & 15/2$^{-}$  & 11/2$^{-}$   & 1.92(6)  & 0.08(3)  & 0.38(16)  & E2 \\
4076.8(2)& 292.6(1)  & 3784.3 & 61(1)   & 17/2$^{-}$  & 15/2$^{-}$   & 1.15(5)  & -0.28(7) & -0.71(18) & M1(+E2) \\
         & 835.2(1)  & 2716.9 & 318(3)  & 17/2$^{-}$  & 13/2$^{-}$   & 1.75(4)  & 0.15(4)  & 0.59(14)  & E2 \\
4181.1(2)& 640.3(1)  & 3540.8 & 41(1)   & 15/2        & 13/2         & 0.97(9)  &          &           & D   \\
4287.4(2)& 566.7(1)  & 3720.7 & 32(1)   & 19/2        & 15/2         & 1.86(15) &          &           & Q   \\
4475.8(2)& 843.4(1)  & 3632.4 & 78(1)   & 19/2$^{-}$  & 17/2$^{+}$   & 0.91(5)  & 0.25(9)  & 0.96(34)  & E1 \\
4526.3(2)& 449.5(1)  & 4076.8 & 254(3)  & 21/2$^{-}$  & 17/2$^{-}$   & 2.08(7)  & 0.19(4)  & 0.53(11)  & E2 \\
4800.4(2)& 1168.1(1) & 3632.4 & 205(1)  & 21/2$^{+}$  & 17/2$^{+}$   & 1.76(9)  & 0.12(4)  & 0.54(19)  & E2 \\
4872.6(2)& 1331.8(1) & 3540.8 & 21(1)   &             & 13/2         &          &          &           &  \\
5480.0(2)& 1004.1(1) & 4475.8 & 29(1)   & 21/2        & 19/2$^{-}$   & 1.17(11) &          &           & D+Q  \\
5677.0(2)& 1150.7(1) & 4526.3 & 24(1)   & 23/2        & 21/2$^{-}$   & 1.58(14) &          &           & D+Q \\
5729.9(2)& 1203.6(1) & 4526.3 & 120(2)  & 25/2$^{-}$  & 21/2$^{-}$   & 1.81(17) & 0.11(3)  & 0.54(15)  & E2 \\
5978.3(2)& 498.2(1)  & 5480.0 & 34(1)   & 25/2$^{+}$  & 21/2         & 2.09(15) &          &           & Q \\
         & 1177.9(1) & 4800.4 & 146(1)  & 25/2$^{+}$  & 21/2$^{+}$   & 1.81(10) & 0.12(5)  & 0.57(21)  & E2 \\
6077.5(2)& 1277.1(1) & 4800.4 & 23(1)   & 23/2        & 21/2$^{+}$   & 0.72(23) &          &           & D+Q \\
7105.4(2)& 1127.1(1) & 5978.3 & 118(1)  & 29/2$^{+}$  & 25/2$^{+}$   & 1.81(17) & 0.05(5)  & 0.43(23)  & E2 \\
8343.3(2)& 1237.9(1) & 7105.4 & 96(1)   & 33/2$^{+}$  & 29/2$^{+}$   & 1.75(14) & 0.07(5)  & 0.34(25)  & E2  \\

\hline
\bigskip
\end{longtable*}

It is noteworthy that $^{69}$Ga wasn't one of the dominantly populated nuclei in the experiment (albeit with sufficient evidence 
of new information to provide impetus for this study). Consequently, a number 
of newly identified transitions did not have sufficient count statistics to facilitate determination of the 
respective polarization values, in particular, and, at certain instances, the $R_{ADO}$ values.
In the absence of the polarization measurement, the electric/ magnetic nature of the respective transition
could not be ascertained and the parity of its initial state could not be assigned. The multipolarities 
of such transitions, for which only the $R_{ADO}$ could be determined, have been accordingly assigned (Table~1) 
as D (dipole) or Q (quadrupole) or D+Q (mixed), as per the respective $R_{ADO}$ value. The list of    
these $\gamma$-rays include the newly identified 1004-, 640-, 153-, 1277-, 567-, 1151- and 208-keV.
The low energies of the 153- and the 208-keV transitions limit their polarization measurement
that is based on (Compton) scattering between adjacent crystals (of the HPGe clover detector) and the 
probability of the same is known to be modest at lower energies ($\lessapprox$ 300-keV).
The remaining transitions in the aforementioned list have been observed to be of low intensities, $\approx$ 3\% of the ground state transition,
that made it obscure to extract reliable results for their polarization assignments. 
Two transitions, 1106-keV de-exciting the 1106-keV state and 1235-keV de-exciting the 2571-keV state, had been known 
previously \cite{Bak82} along with spin-parities of the respective levels. However, owing to the  
mere intensities ($\lessapprox$ 3-4\% of the ground state transition) of these transitions, their polarization could not be 
measured in the present study and, consequently, the (known) parities of the de-excited levels could not be confirmed/ validated
herein. Following the $R_{ADO}$ ratios in the current analysis, the two transitions have been assigned mixed multipolarity D+Q, while 
their electric/ magnetic nature and the parities of the respective states have not been assigned. 
Polarization asymmetry could not be determined also for the newly identified $\gamma$-rays 498-keV and 1067-keV respectively 
de-exciting the previously known levels 5978-keV and 3784-keV. However, these transitions represent only $\lessapprox$ 20\% of the
decay of the respective levels and the previously \cite{Bak82} assigned spin-parities of these states have been validated 
from the $R_{ADO}$ and polarization values of the stronger branches de-exciting them; that is the 1178-keV transition for the 5978-keV level 
and the 1118-keV transition for the 3784-keV level. As far as the 1332-keV transition is concerned, it has been newly identified 
in this work but owing to its low intensity, neither $R_{ADO}$ nor polarization could be analyzed and the respective level at 4873-keV 
remains without any spin-parity assignment. Even the two previously known transitions, 1146- and 1353-keV, de-exciting the 
3117-keV state, were weakly observed in this data and neither the $R_{ADO}$ nor the polarization could be determined 
for these $\gamma$-rays. Consequently, the previously assigned spin-parity of the the 3117-keV level could not be 
validated in the present analysis and has thus not been included in the results (level scheme and table). \\

The newly identified 1971-keV transition is of particular significance. It connects the 9/2$^+_1$ state at 1971-keV and the 3/2$^-_1$ ground state.
The spin-parities of the initial and the final state of this transition are thus otherwise established and the absence of $R_{ADO}$ and polarization 
measurements for this weak transition do not come in the way of its E3 assignment (Table~I). It is interesting to note that a similar transition (2074-keV, 9/2$^+$$\rightarrow$$3/2^-$)
has also been reported \cite{Zho24} in the excitation scheme of $^{67}$Ga and might be interpreted as a possible signature of octupole correlations 
in these isotopes. \\

The gaps in the assignments notwithstanding, this maiden endeavor for high resolution nuclear structure investigation of $^{69}$Ga nucleus
populated in a heavy-ion induced reaction, has substantially extended its excitation scheme and the interpretation of the 
same is detailed in the subsequent section.

\section{Discussions}

\begin{figure*}
\includegraphics[angle=0,scale=0.37,trim=0.0cm 0.6cm 1.5cm 1.5cm,clip=true]{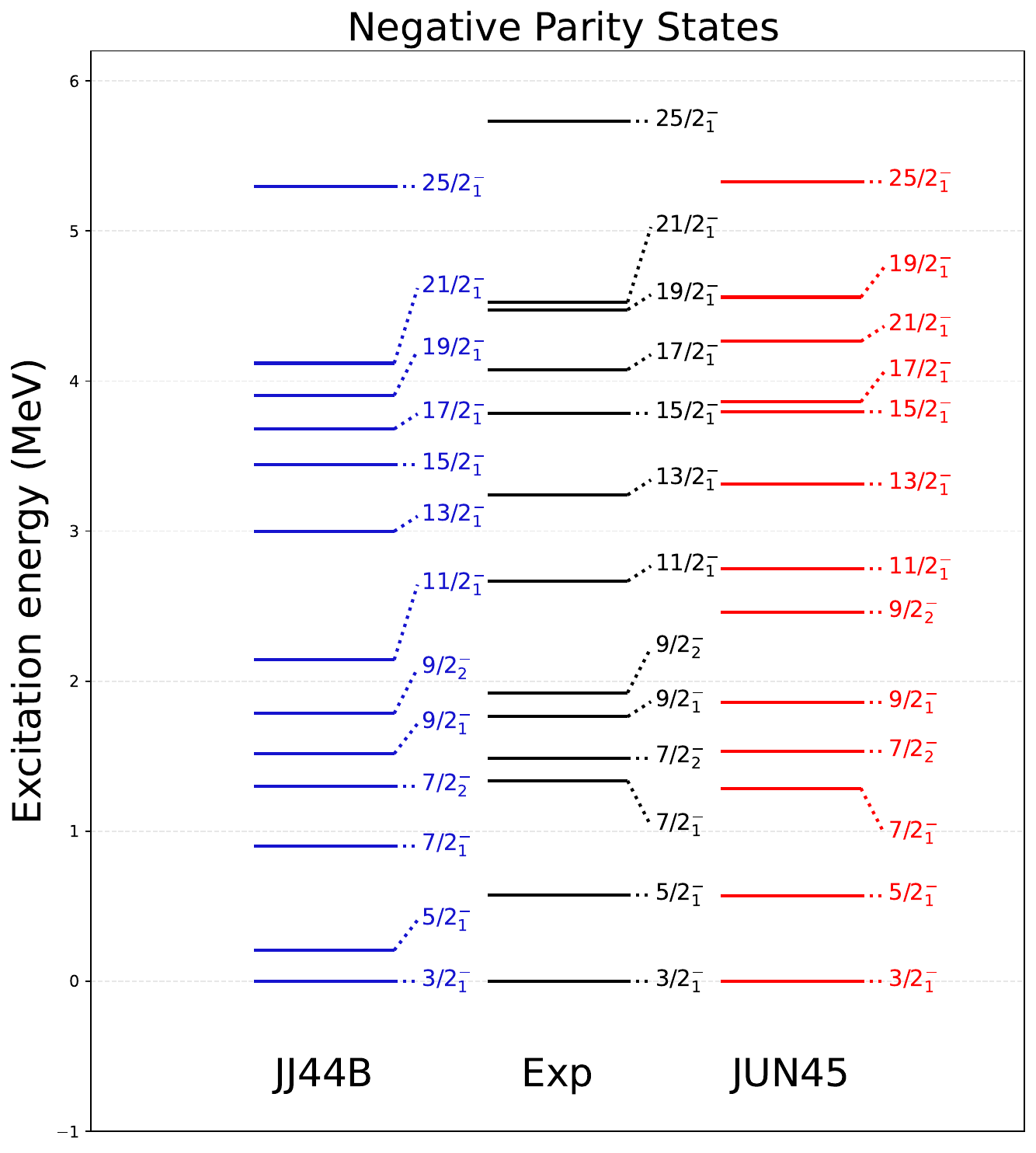}
\includegraphics[angle=0,scale=0.37,trim=0.0cm 0.6cm 1.5cm 2.0cm,clip=true]{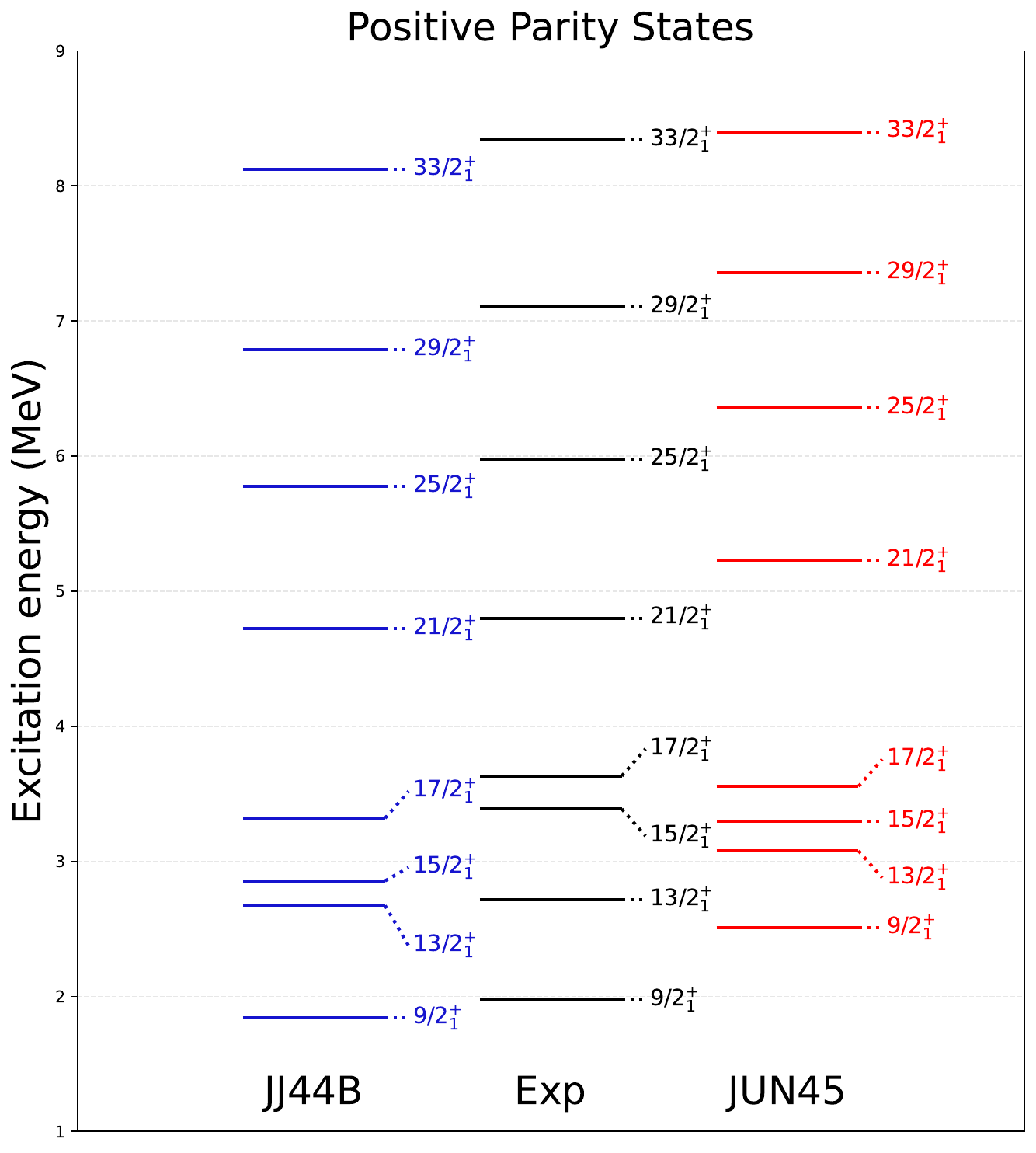}
\caption{\label{fig7} Comparison between experimental and calculated level energies of $^{69}$Ga. The respective interaction, JUN45 and JJ44B, used in the 
calculations are labeled accordingly.}
\end{figure*}

One of the primary objectives of this work has been to interpret the observed excitation scheme of $^{69}$Ga ($Z = 31, N =38$) nucleus within the framework of 
large basis shell model calculations and, in the process, validate the application of the latter in the context of structure studies of transitional nuclei
with considerable (13 in this case) number of nucleons in the valence space outside the core. Large basis shell model calculations have been carried out 
using the KSHELL \cite{Shi19} code. The chosen model space was one consisting of $p_{3/2}, f_{5/2}, p_{1/2}, g_{9/2}$ orbitals outside the doubly-magic $^{56}$Ni ($Z, N = 28$) core.
The three valence protons and the ten valence neutrons of the $^{69}$Ga nucleus, outside the Ni-core, were allowed unrestricted occupancy of the orbitals 
in the aforementioned model space. Two different interactions, JJ44B \cite{Lis04} and JUN45 \cite{Hon09}, were used in the calculations. 
The single particle energies used in the calculations with JUN45 are $p_{3/2}$: 9.828 MeV, $f_{5/2}$: 8.709 MeV, $p_{1/2}$: 7.839 MeV, $g_{9/2}$: 6.262 MeV.
The same for J44B interaction are, $p_{3/2}$: 9.657 MeV, $f_{5/2}$: 9.286 MeV, $p_{1/2}$: 8.270 MeV, $g_{9/2}$: 5.894 MeV.
JUN45 is known to be 
extensively applied for nuclei in the immediate \cite{Sam19} vicinity of the $Z, N =28$ closure as well as for those with considerable \cite{Kay25} number of nucleons
in the $f_{5/2}pg_{9/2}$ valence space. Both JUN45 and JJ44B interactions were used by Zhong {\it{et al.}} \cite{Zho21} for the next heavier (and neutron richer) isotope $^{71}$Ga ($N = 40$).
They reported modest or poor overlap between the experimental and the calculated level energies for most of the negative parity states while the 
theoretical energies for positive parity levels were in better agreement, particularly those corresponding to the JUN45 interaction. \\

\begin{longtable*}{cc|ccccccc|ccccccc}
\caption{\label{tab1} Experimental (rounded off to the nearest integer) and calculated level energies of $^{69}$Ga along with the dominant single particle configurations of the states. 
The contributions of the dominant configurations are expressed in \%, rounded off to the nearest integer. 
The calculations have been carried out using two different interactions, JUN45 and JJ44B, and the respective results are accordingly recorded in the table.} \\
\hline
Exp (keV) & $J^{\pi}$ & JUN45 (keV) & & $p_{3/2}$ & $f_{5/2}$ & $p_{1/2}$ & $g_{9/2}$ & \% & JJ44B & & $p_{3/2}$ & $f_{5/2}$ & $p_{1/2}$ & $g_{9/2}$ & \% \\
\hline
\endfirsthead

\multicolumn{16}{c}{{\tablename\ \thetable{} -- continued from previous page}} \\
\hline
Exp (keV) & $J^{\pi}$ & JUN45 (keV) & & $p_{3/2}$ & $f_{5/2}$ & $p_{1/2}$ & $g_{9/2}$ & \% & JJ44B & & $p_{3/2}$ & $f_{5/2}$ & $p_{1/2}$ & $g_{9/2}$ & \% \\
\hline
\endhead

\hline \multicolumn{16}{c}{{continued on next page}} \\
\endfoot

\hline
\endlastfoot
          &                   &      &   &      &      &      &      &      &      &   &      &      &      &      &    \\
           \multicolumn{14}{c}{\bf{Negative Parity States}} \\
          &                   &      &   &      &      &      &      &      &      &   &      &      &      &      &    \\
        0 & 3/2$^{-}_{1}$     & 0    & p & 3    & 0    & 0    & 0    &  32  & 0    & p & 1,3  & 0,2  & 0    & 0    & 25 \\
          &                   &      & n & 2,4  & 2,4,6& 0,2  & 0    &      &      & n & 2,3,4& 2,4  & 0,1,2& 0,2  &  \\
      574 & 5/2$^{-}_{1}$     & 571  & p & 2    & 1    & 0    & 0    &  31  & 208  & p & 2    & 1    & 0    & 0    & 27 \\
          &                   &      & n & 2,4  & 2,4,6& 0,2  & 2    &      &      & n & 2,3,4& 2,4  & 0,1,2& 2    &  \\
     1336 & 7/2$^{-}_{1}$     & 1286 & p & 3    & 0    & 0    & 0    &  35  & 902  & p & 1,2  & 1,2  & 0    & 0    & 27 \\
          &                   &      & n & 3,4  &   4,5& 1,2  & 0    &      &      & n & 2,3,4& 4    & 0,1,2& 2    &   \\
     1487 & 7/2$^{-}_{2}$     & 1531 & p & 2    & 1    & 0    & 0    &  31  & 1302 & p & 1,2  & 1,2  & 0    & 0    & 22\\
	  &                   &      & n & 2,3,4& 2,3,4& 0,1,2& 2    &      &      & n & 2,3,4& 3,4  & 0,1,2& 2    &   \\
     1765 & 9/2$^{-}_{1}$     & 1857 & p & 2    & 1    & 0    & 0    &  38  & 1517 & p &   2  & 1    & 0    & 0    & 28\\
          &                   &      & n & 2,3,4&2,3,4 & 0,1,2& 2    &      &      & n & 2,3,4& 2,3,4& 0,1,2& 2    &    \\
     1923 & 9/2$^{-}_{2}$     & 2459 & p & 2    & 1    & 0    & 0    &  31  & 1786 & p & 1    & 2    & 0    & 0    & 20 \\
          &                   &      & n & 2,3,4& 3,4,5& 0,1,2& 2    &      &      & n & 2,3,4&   3,4& 0,1,2& 2    &     \\
     2667 & 11/2$^{-}_{1}$    & 2749 & p & 3    & 0    & 0    & 0    &  46  & 2143 & p & 1,2  & 1,2  & 0,1  & 0    & 30  \\
          &                   &      & n & 3,4  & 4,5  & 2,0  & 0    &      &      & n & 2,3,4&   4  & 0,1,2& 2    &      \\
     3241 & 13/2$^{-}_{1}$    & 3312 & p & 2    &   1  & 0    & 0    &  37  & 2999 & p &   2  & 1    & 0    & 0    & 26   \\
	  &                   &      & n & 2,3,4& 2,3,4& 0,1,2& 2    &      &      & n & 2,3,4& 3,4  & 0,1,2& 2    &      \\
     3784 & 15/2$^{-}_{1}$    & 3795 & p &   2  & 0,1  & 0,1  & 0    &  26  & 3441 & p & 1,2  & 1,2  & 0,1  & 0    & 46    \\
	  &                   &      & n & 2,3,4& 3,4  & 0,1,2& 2    &      &      & n & 2,3,4& 3,4  &0,1,2 & 2    &      \\
     4077 & 17/2$^{-}_{1}$    & 3862 & p &   2  &   1  & 0    & 0    &  31  & 3681 & p &   2  & 1    & 0    & 0    & 21    \\
	  &                   &      & n & 2,3,4& 3,4  & 0,1,2& 2    &      &      & n & 2,3,4& 4    & 0,1,2& 2    &        \\
     4476 & 19/2$^{-}_{1}$    & 4559 & p & 1,2,3& 0,2  & 0,1  & 0    &  25  & 3905 & p & 1    & 1,2  & 0,1  & 0    & 27     \\
	  &                   &      & n & 3,4  & 3,4  & 0,1  & 2    &      &      & n & 2,3,4& 4    & 0,1,2& 2    &        \\
     4526 & 21/2$^{-}_{1}$    & 4265 & p & 2    & 1    & 0    & 0    &  46  & 4119 & p &   2  & 1    & 0    & 0    & 36      \\
	  &                   &      & n & 2,3,4& 2,3,4& 0,1,2& 2    &      &      & n & 2,3,4& 2,4  & 0,1  & 2    &         \\
     5730 & 25/2$^{-}_{1}$    & 5328 & p & 2    & 1    & 0    & 0    &  55  & 5297 & p &   2  & 1    & 0    & 0    &  45     \\
          &                   &      & n & 2,3,4& 3,4  & 0,1,2& 2    &      &      & n & 3,4  & 3,4  & 0,1,2& 2    &         \\
	  &                   &      &   &      &      &      &      &      &      &   &      &      &      &      &    \\
	   \multicolumn{14}{c}{\bf{Positive Parity States}} \\
	  &                   &      &   &      &      &      &      &      &      &   &      &      &      &      &    \\
     1972 & 9/2$^{+}_{1}$     & 2507 & p & 2,3  & 0    & 0,1  & 0,1  &  24  & 1839 & p & 0    & 2    & 0    & 1    & 18    \\
	  &                   &      & n & 4    & 4,5  & 0,1  & 1,2  &      &      & n & 2,3,4& 4    & 0,1,2& 2    &     \\
     2717 & 13/2$^{+}_{1}$    & 3078 & p & 2,3  & 0,1  & 0,1  & 0    &  47  & 2673 & p & 1    & 2    & 0    & 0    & 39  \\
	  &                   &      & n & 4    & 3,4,5& 0,1,2& 1    &      &      & n & 2,3,4& 3,4  & 0,1,2& 3    &     \\
     3388 & 15/2$^{+}_{1}$    & 3295 & p & 2    & 1    & 0    & 0    &  38  & 2855 & p & 1,2  & 1,2  & 0,1  & 0    & 39   \\
	  &                   &      & n & 2,3,4& 3,4,5& 0,1  & 1,3  &      &      & n & 2,3,4& 3    & 0,1,2& 3    &       \\
     3632 & 17/2$^{+}_{1}$    & 3555 & p & 2    & 1    & 0    & 0    &  52  & 3319 & p & 2    & 1    & 0    & 0    &  33   \\
          &                   &      & n & 2,3,4& 3,4,5& 0,1,2& 1    &      &      & n & 2,3,4&  3,4 & 0,1,2& 1,3  &      \\
     4800 & 21/2$^{+}_{1}$    & 5229 & p & 2    & 1    & 0    & 0    &  45  & 4722 & p & 1,2  & 1,2  & 0,1  & 0    & 44   \\
          &                   &      & n & 2,3,4& 3,4,5& 0,1,2& 1    &      &      & n & 2,3,4& 3    & 0,1,2& 3    &      \\
     5978 & 25/2$^{+}_{1}$    & 6355 & p & 2    & 0    & 0    & 1    &  34  & 5776 & p & 0,1,2& 0,1,2& 0,1  & 1    & 42 \\
	  &                   &      & n & 2,3,4& 2,3,4& 0,1,2& 2    &      &      & n & 2,3,4&     4& 0,1,2& 2    &     \\
     7105 & 29/2$^{+}_{1}$    & 7357 & p & 0,1,2& 0,1,2& 0,1  & 1    &  44  & 6789 & p & 0,1,2& 0,1,2& 0,1  & 1    & 52  \\
          &                   &      & n & 2,3,4& 3,4  & 0,1,2& 2    &      &      & n & 3,4  & 3,4  & 0,1  & 2    &     \\
     8343 & 33/2$^{+}_{1}$    & 8401 & p & 1    & 1    & 0    & 1    &  45  & 8123 & p & 0,1  & 1    & 0,1  & 1    & 58   \\
          &                   &      & n & 2,3,4& 2,3,4& 0,1,2& 2    &      &      & n & 2,3,4& 2,3,4& 0,1,2& 2    &       \\

\end{longtable*}

\begin{figure}
\includegraphics[angle=0,scale=.28,trim=0.0cm 0.0cm 0.0cm 0.0cm,clip=true]{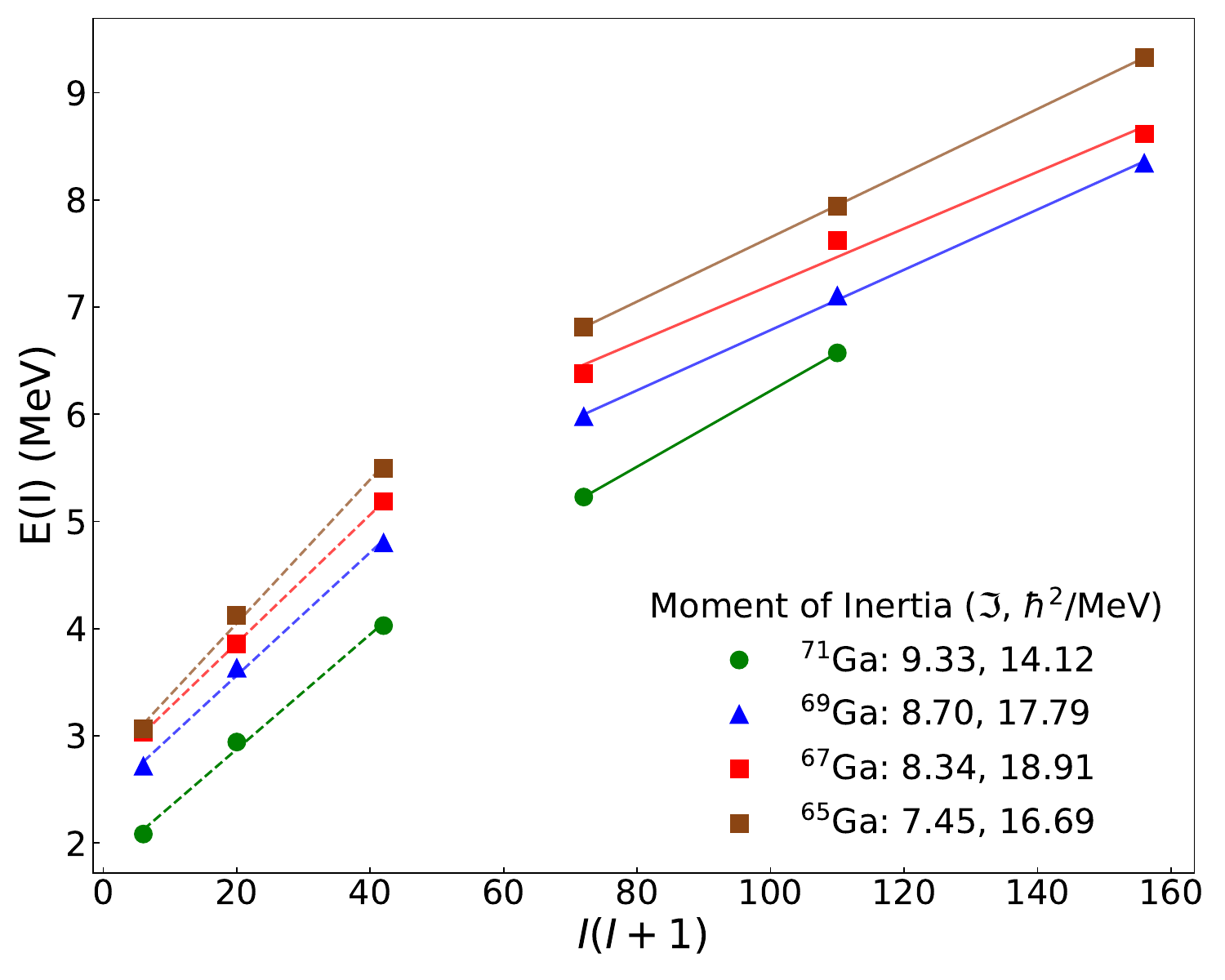}
\includegraphics[angle=0,scale=.33,trim=1.0cm 1.0cm 1.0cm 2.0cm,clip=true]{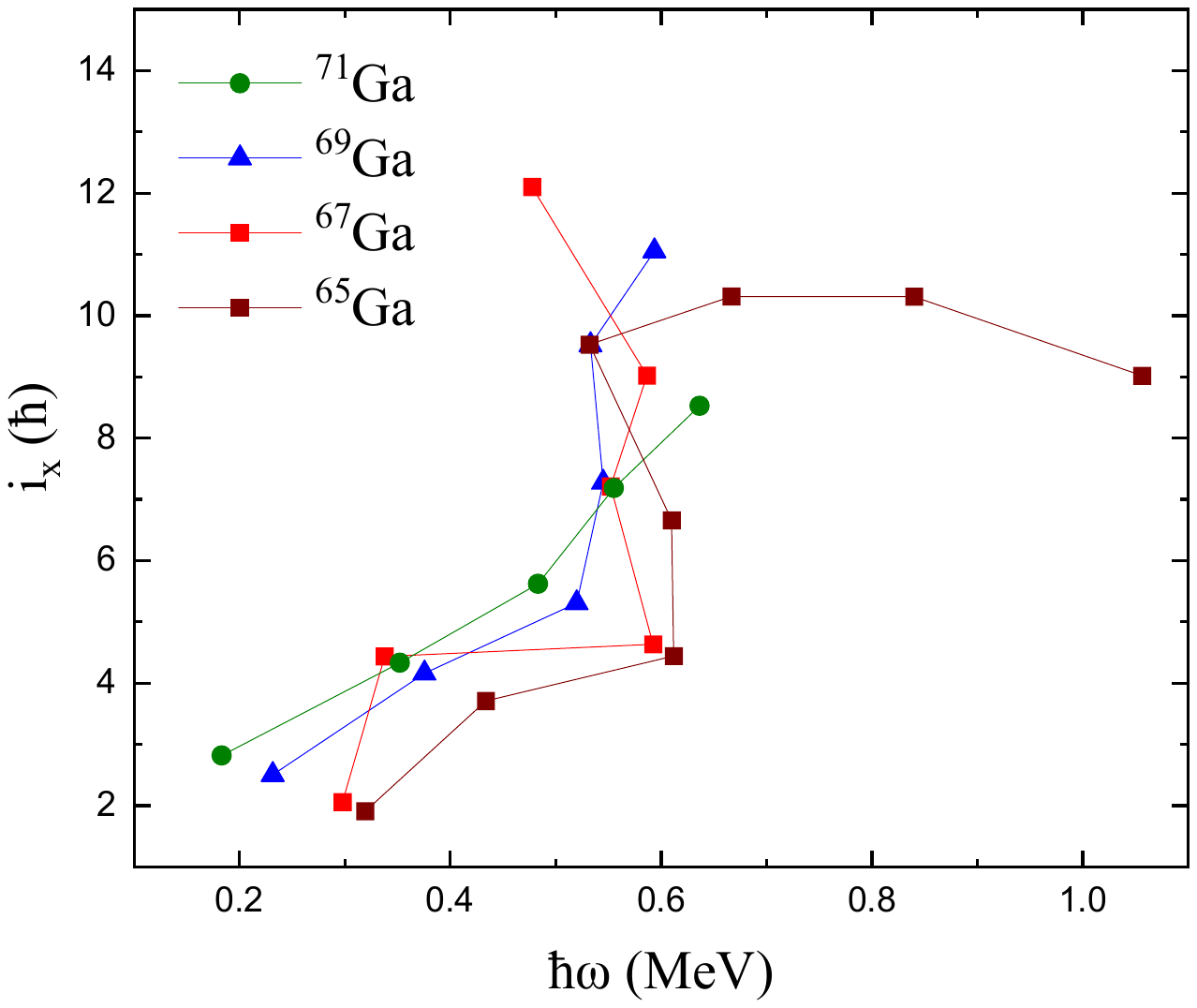}
\caption{\label{fig8}Plot of (a) level energy ($E(I)$) vs $I(I+1)$ and  
(b) aligned angular momentum ($i_x$) vs rotational frequency ($\hbar\omega$) for the $\pi g_{9/2}$ band in $^{69}$Ga and its neighboring isotopes. 
The Moment of Inertia ($\Im$) values resulting from the fit of $E(I)$ vs $I(I+1)$ data, before and after alignment and using $E(I) = E_0 + (\hbar^2)/2\Im)(I(I+1))$, are included in the respective plot.}
\end{figure}

The results of the shell model calculations, carried out in this study, are recorded in Table~II and Fig.~7.
Table~II lists the calculated level energies corresponding to the two interactions vis-a-vis the experimental ones,
along with the dominant particle configurations of the respective states.
The configurations are generally mixed ones, and the representative occupancies listed in Table~II are drawn from individual partitions 
that contribute at least 5\% to the total wavefunction. 
The 3/2$^-$ ground state is reproduced by both JUN45 and JJ44B interactions, albeit the dominant particle configurations 
for the latter include those at variance from the known/ expected $\pi p_{3/2}^3$ occupancy. 
The calculated level energies of the negative parity states following the JUN45 interaction are overall in superior agreement (Table~II and Fig.~7) 
with their experimental values; the
difference being mostly within $\approx$ 100-keV. Deviations in this trend 
occur for 9/2$^-$ yrare state as well as for the highest negative parity states observed at 21/2 and 25/2. 
The calculated energy of the 25/2$^-$ level corresponding to the JJ44B interaction is also considerably deviant 
from the measurements. In fact, the agreement between the JJ44B calculated energies and the experimental ones
is only modest, for all the negative parity states, except for the 9/2$^-$ yrare one.
The particle configurations of these levels, corresponding to the calculations with JUN45,    
are dominantly based on $\pi p_{3/2}^{2,3}f_{5/2}^{0,1}$ partitions, for those at lower excitations, and includes 
contributions of $\pi p_{3/2}^{1-3}f_{5/2}^{0,1}p_{1/2}^{0,1}$ partitions, for negative parity levels above 11/2$\hbar$.   
It is, however, noteworthy that the JUN45 calculations exclude the $\pi p_{1/2}$ occupancy in the major partitions 
associated with the highest odd-parity states, 21/2 and 25/2; as noted from the table, the calculated energies 
for these levels are also at considerable variance with the experimental ones. The neutron occupancies underlying the
negative parity states, in both (JUN45 and JJ44B) calculations, interestingly includes considerable contributions 
from partitions with $\nu g_{9/2}^{2}$ that also consist of even (or zero) number of neutrons in each of $p_{3/2}f_{5/2}p_{1/2}$ orbitals
or odd number in two of them. These particle configurations, particularly in the context of the neutron occupancies, 
are in overlap with those reported for the $^{71}$Ga \cite{Zho21} nucleus. The proton occupancies corresponding to the dominant partitions, however,  
do not include $\pi p_{1/2}$ and the general agreement between the calculated and the experimental level energies, for the negative parity states,
is rather modest therein. \\

The present calculations manifest somewhat sporadic overlap between the calculated and the experimental level energies, for the positive parity states.
The lowest ones (9/2$^+$ and 13/2$^+$) are very well reproduced in the JJ44B calculations while the next higher ones (15/2$^+$ and 17/2$^+$) exhibit 
superior agreement for JUN45. The next higher state, 21/2$^+$, again has its energy, calculated with the JJ44B interaction, close to the experimental value while for
the 25/2$^+$ state, the agreement is modestly better for JJ44B vis-a-vis JUN45, the difference (between the calculated and the experimental energies) 
being respectively $\approx$ 200 and $\approx$ 400-keV. The calculated energy of the 29/2$^+$ state deviate differently (lesser and greater by $\approx$ 250-300 keV) 
in the two calculations while the highest positive parity state, 33/2$^+$, is excellently reproduced by JUN45. It is conjectured that the 
varying overlaps across the positive parity states, between the two calculations, might be sourced in the microscopic details of the respective interactions
that still require refinements towards constraining the calculations and the same is expected to be facilitated by the availability of 
more stringent data, such as on the level lifetimes and transition probabilities. The particle configurations for the positive parity states, of low and intermediate excitations,
are majorly (Table~II) contributed by partitions with $\nu g_{9/2}^3$. The three protons are distributed in the $f5p$ orbitals and 
there is no proton occupancy of the $g_{9/2}$ orbital significantly contributing to the corresponding wavefunctions; 9/2$^+$ being an exception. The higher positive parity levels, on the contrary, are 
dominantly contributed by $\pi g_{9/2}^1$ occupancy along with a neutron-pair in the $g_{9/2}$ orbital. These configurations are at variance to those reported \cite{Zho21} for the
(same) positive parity states in $^{71}$Ga. The calculated energies in the latter are widely discrepant from the experimental ones and the associated particle configurations 
are dominated by partitions corresponding to no proton occupancy of the $g_{9/2}$ orbital and neutron configurations $\nu g_{9/2}^1$ for the states at lower excitations 
and $\nu g_{9/2}^3$ for the higher (positive parity) ones. \\

\begin{figure}
\includegraphics[angle=90,scale=.30,trim=0.0cm 0.0cm 4.0cm 0.0cm,clip=true]{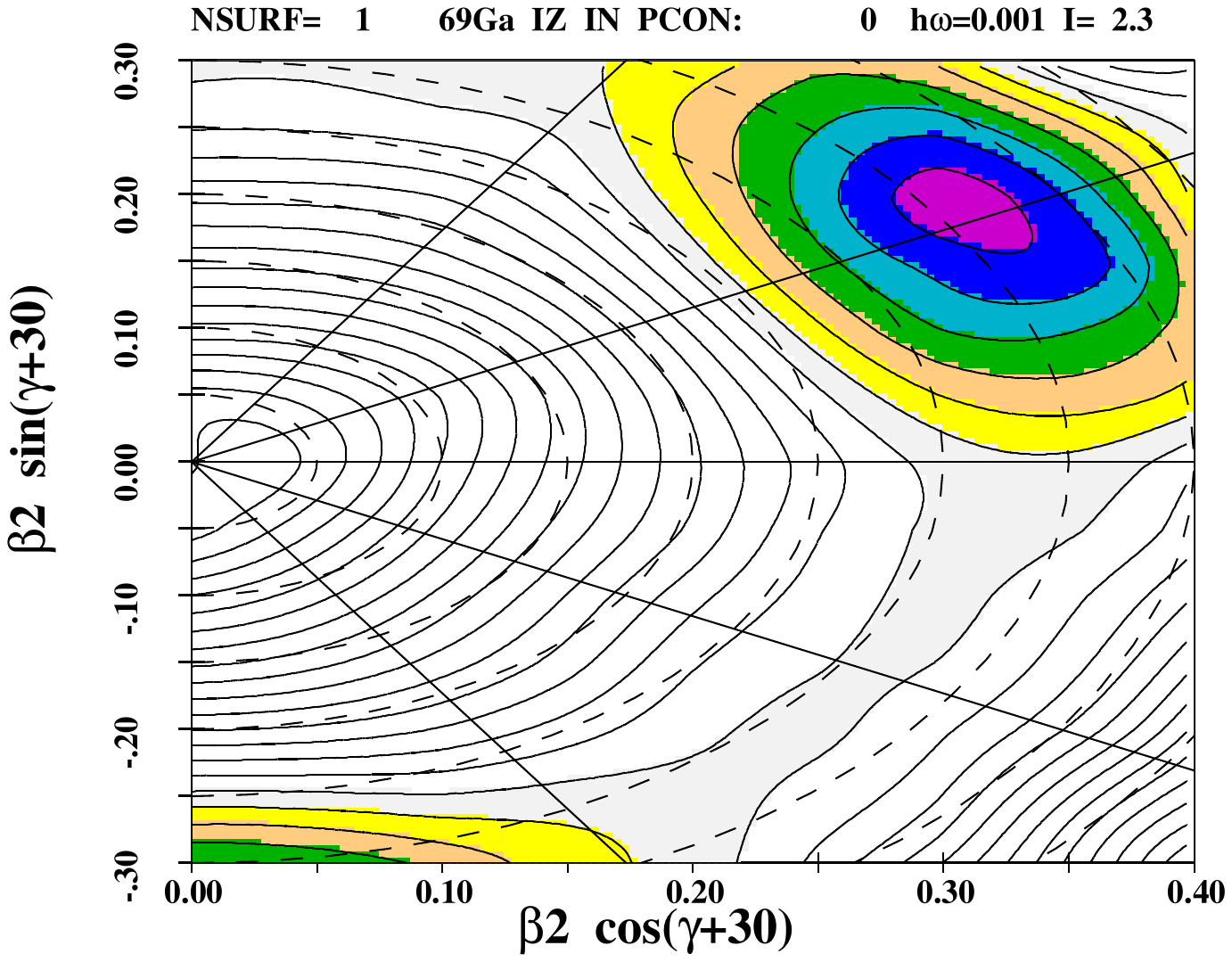}
\includegraphics[angle=90,scale=.30,trim=0.0cm 0.0cm 4.0cm 0.0cm,clip=true]{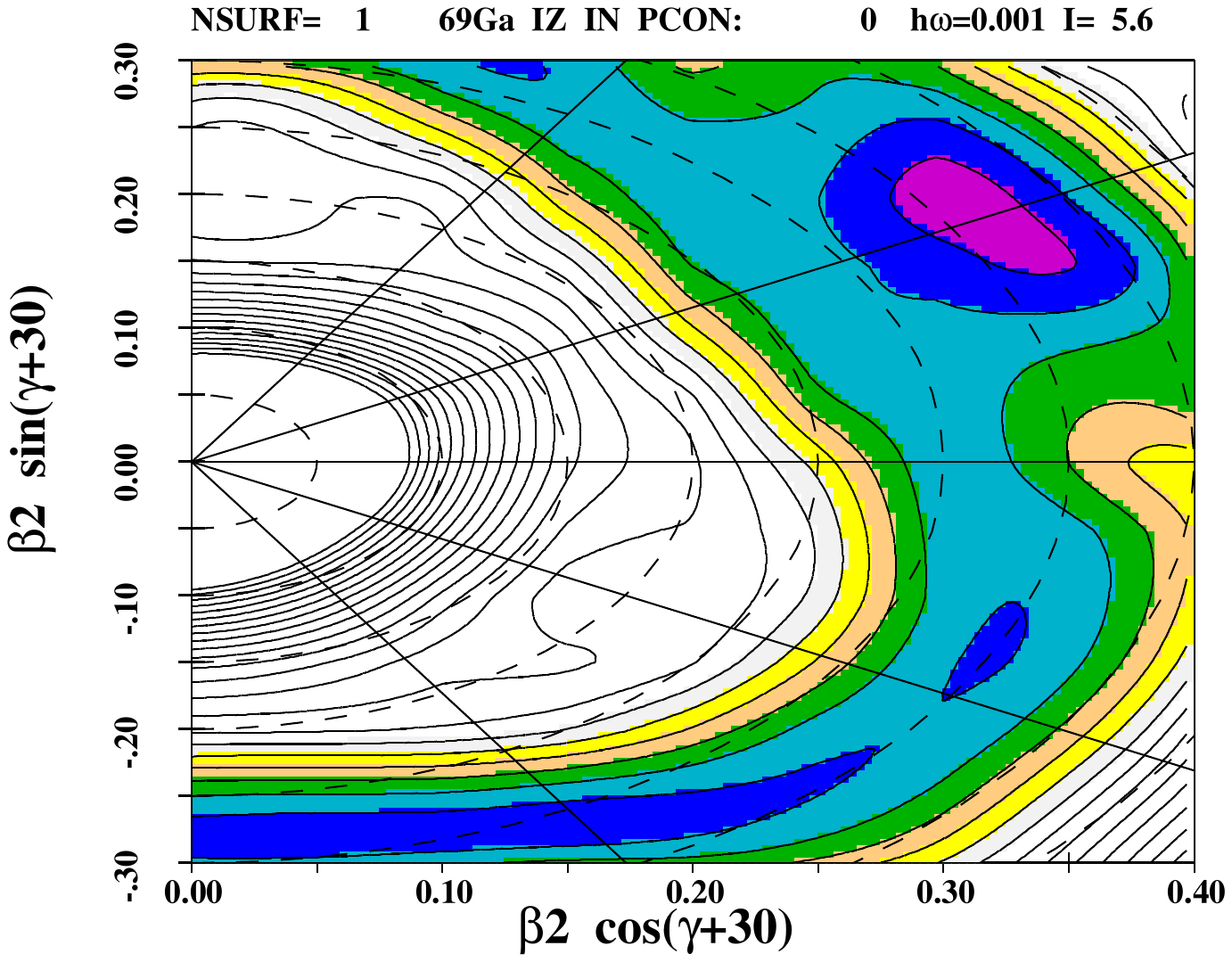}
\caption{\label{fig9} TRS plots of the 9/2$^+$ band corresponding to configurations $\pi g_{9/2}^1$ before (upper panel) and $\pi g_{9/2}^1 \otimes \nu g_{9/2}^2$ after (lower panel) the 
neutron-pair alignment, at near minimum rotational frequency ($\hbar\omega$) for each configuration.}
\end{figure}

The yrast 9/2$^+$, 13/2$^+$, 17/2$^+$, 21/2$^+$, 25/2$^+$, 29/2$^+$, 33/2$^+$ states in $^{69}$Ga constitute a band that is similar to the sequence 
observed \cite{Wei01,Zho24,Zho21} in the neighboring isotopes of the nucleus. It is understood the sequence is based on $\pi g_{9/2}$ that is 
at variance with the dominant particle configurations of most of the low-lying states of the sequence as emerging from the present shell model calculations or, for instance, those
reported \cite{Zho21} for the $^{71}$Ga nucleus. 
The 9/2$^+_1$ state of $^{69}$Ga that is calculated using the JJ44B interaction, however, corresponds to the $\pi g_{9/2}^1$ configuration and the respective level energy overlaps excellently with the
experimental value. This calculated configuration of the 9/2$^+$ band-head is but an isolated instance while the calculations for the other (13/2$^+$, 17/2$^+$, 21/2$^+$) states at low excitations 
do not bring out any dominance of the $\pi g_{9/2}^1$ occupancy. 
The band has been known \cite{Wei01} to exhibit alignment of $g_{9/2}$ neutron pair in other isotopes of Ga 
and extend to high spins following the gain in the angular momentum. 
However, the major single particle occupancies indicated by shell model calculations do not conclusively indicate such neutron excitations in the $g_{9/2}$ orbital, 
particularly associated with any change in the configuration (alignment) within the sequence. 
Since this band in $^{69}$Ga has been majorly extended in this study, it is required that its 
characteristics be explored and compared with those of the equivalent structures in other isotopes. Fig.~8 illustrates the different aspects of the $\pi g_{9/2}$ positive
parity band in $^{69}$Ga vis-a-vis those in other isotopes. The fit of excitation energy ($E(I)$) vs $I(I+1)$ data (Fig.~8a) using $E(I) = E_0 + (\hbar^2/2\Im)I(I+1)$, before 
and after the known or envisaged alignments, yield Moment of Inertia ($\Im$) values that are comparable across the isotopic chain, albeit with slight and systematic increase 
of deformation, indicated by $\Im$, with increasing neutron number. The rotational frequency ($\hbar\omega$) corresponding to the $\nu g_{9/2}^2$ alignment correspondingly 
evolves (Fig.~8b) to lower values for heavier isotopes as indicated in the plot (Fig.~8b) of aligned angular momentum ($i_x$) vs $\hbar\omega$, calculated using the 
expressions in Ref.~\cite{Reg03}. Further to the validation of the rotational characteristics of the observed band, it is natural to probe the associated deformation and
the same has been pursued in the present work within the framework of Total Routhian Surface (TRS) calculations. The method is based on Nilsson-Strutinsky formalism 
wherein the single particle states are calculated as a function of the deformation parameters, $\beta_2$ and $\gamma$, in Woods-Saxon potential with global parameters. 
The procedure is detailed in Refs.~\cite{Nay22,Bas23}. The calculated TRS energies are represented as contour plots in $\beta_2$-$\gamma$ plane, after minimization in $\beta_4$, and 
$\gamma$ = 0$^\circ$ (-60$^\circ$) corresponds to the prolate (oblate) shape. The calculations for the positive parity 9/2$^+$ band have been carried out for two configurations, 
$\pi g_{9/2}^1$ and $\pi g_{9/2}^1 \otimes \nu g_{9/2}^2$, in order to probe the associated shape and its changes, if any, following the alignment. The resulting contour plots, 
illustrated in Fig.~9, bring out that the deformation corresponding to the 9/2$^+$ band is strongly prolate at the band head, with $\beta_2$ $\approx$ 0.36, and it  
remains so even after the alignment of the neutron-pair, albeit with indications of onsetting triaxiality and/or $\gamma$-softness. 
The prolate shape is nevertheless considerably stabilized, across the band, presumably owing to the underlying configurations based on the deformation driving
high-$j$ intruder $g_{9/2}$ orbital. \\

It is important to take note of the 1971-keV transition, de-exciting the 9/2$^+$ band-head of the $g_{9/2}$ positive-parity band to the 3/2$^-$ ground state.
As discussed in the previous section, the transition has been assigned an E3 multipolarity, albeit tentatively, and is indicative of possible octupole correlation 
in the $^{69}$Ga nucleus similar to what has been reported for the $^{67}$Ga isotope \cite{Zho24}. The experimental $B(E3)/B(E1)$ value corresponding to $^{69}$Ga is
17.84$\times$10$^7$ fm$^4$, considerably larger than that (4.58$\times$10$^7$ fm$^4$ \cite{Zho24}) in $^{67}$Ga. The increased $B(E3)/B(E1)$ is consistent with the 
proposition of \textquotedblleft strong octupole effects\textquotedblright \cite{Zho24} at $N = 38$ that follows the evolving energy of the 3$-$ state across the Zn ($Z = 30$) 
isotopes and its minimum at $^{68}$Zn (Fig.~6 in Ref.~\cite{Zho24}). \\

\begin{figure}
\includegraphics[angle=0,scale=.25,trim=0.0cm 0.0cm 0.0cm 0.0cm,clip=true]{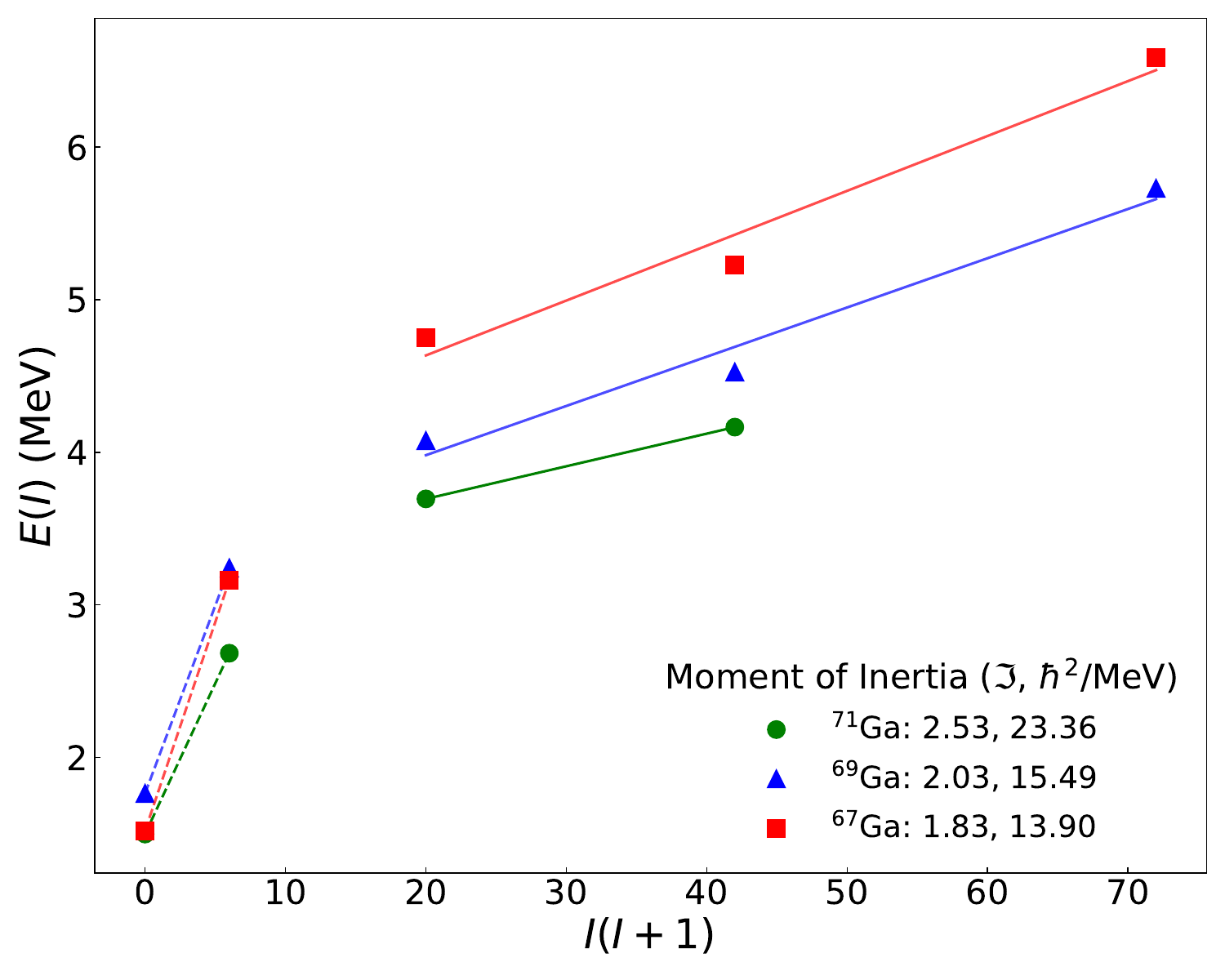}
\includegraphics[angle=0,scale=.30,trim=1.0cm 1.0cm 1.0cm 2.0cm,clip=true]{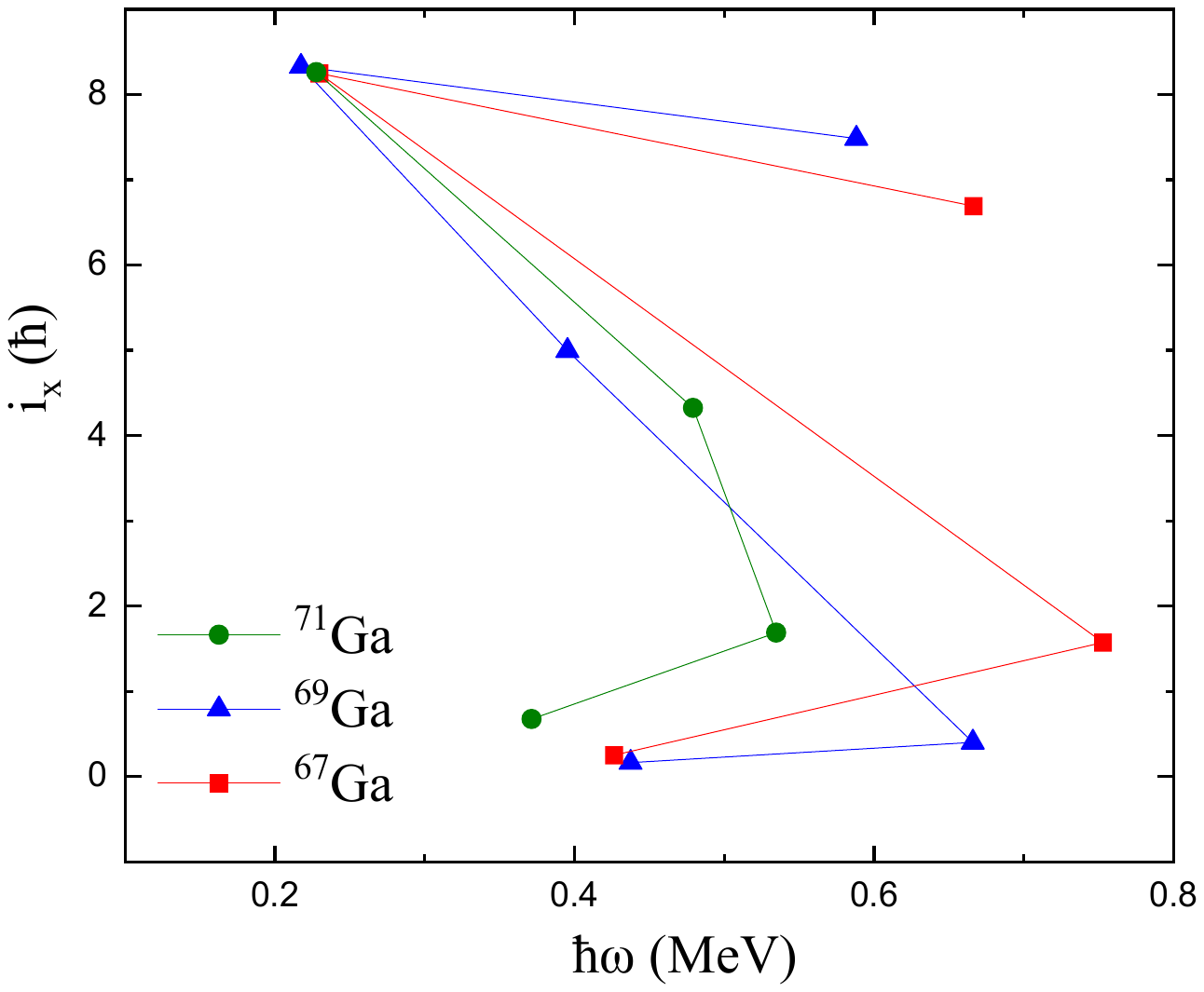}
\caption{\label{fig10}Plot of (a) level energy ($E(I)$) vs $I(I+1)$ and  
(b) aligned angular momentum ($i_x$) vs rotational frequency ($\hbar\omega$) for the $\pi f_{5/2}$ band in $^{69}$Ga and its neighboring isotopes. 
The Moment of Inertia ($\Im$) values resulting from the fit of $E(I)$ vs $I(I+1)$ data, before and after alignment and using $E(I) = E_0 + (\hbar^2)/2\Im)(I(I+1))$, are included in the respective plot.}
\end{figure}

\begin{figure}
\includegraphics[angle=90,scale=.30,trim=0.0cm 0.0cm 4.0cm 0.0cm,clip=true]{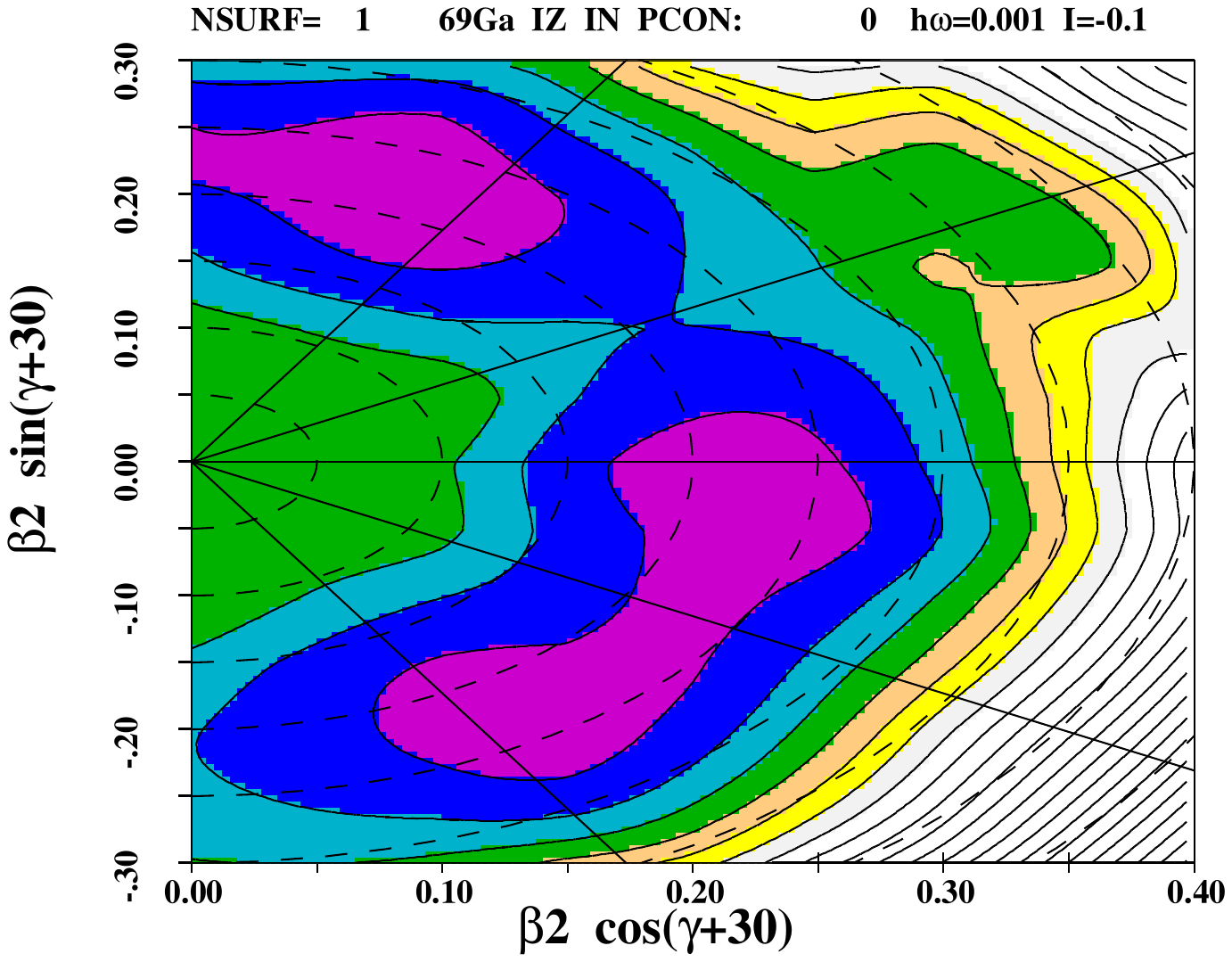}
\includegraphics[angle=90,scale=.30,trim=0.0cm 0.0cm 4.0cm 0.0cm,clip=true]{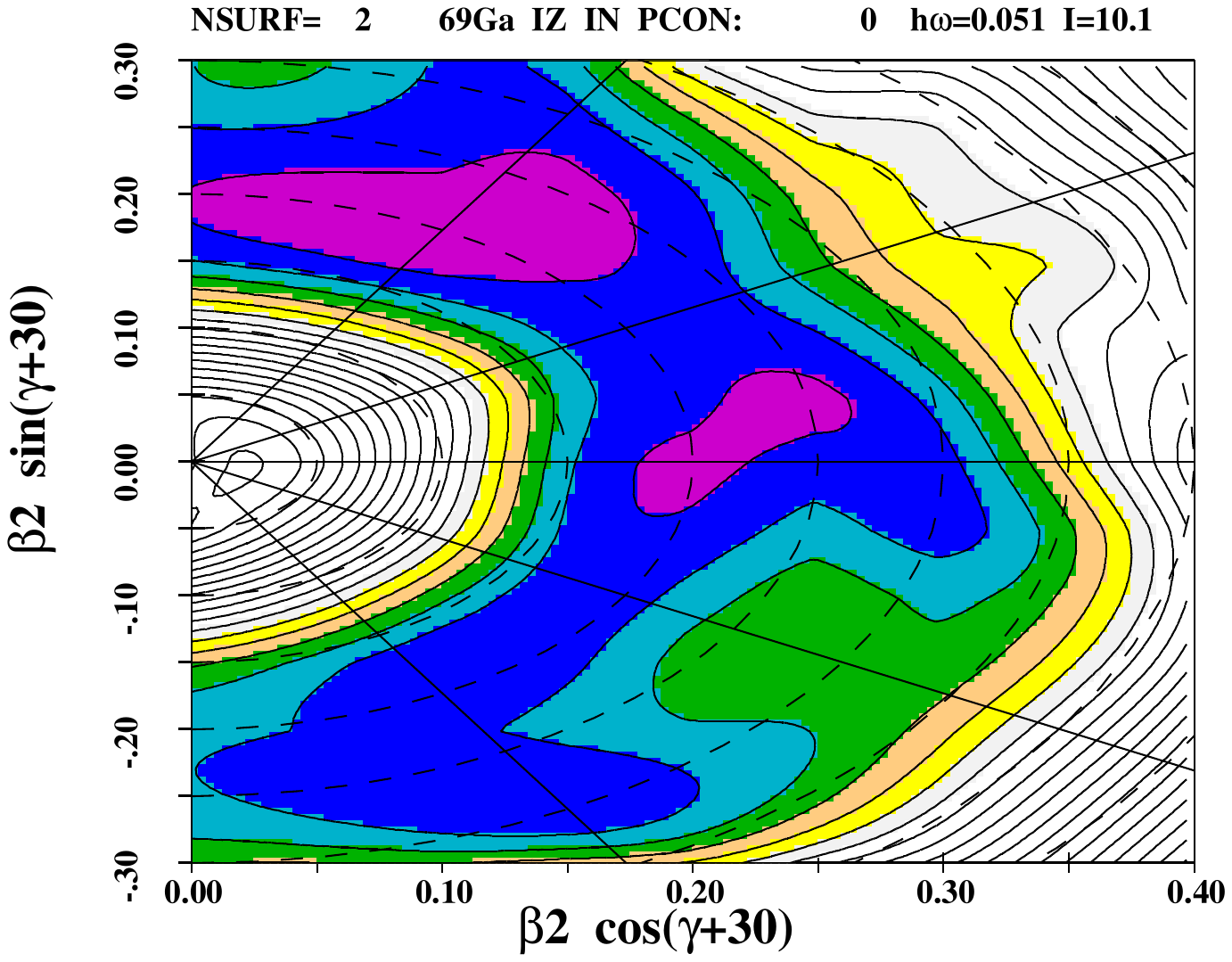}
\caption{\label{fig11} TRS plots for 5/2$^-$ band corresponding to configurations $\pi f_{5/2}^1$ (upper panel) and $\pi f_{5/2}^1 \otimes \nu g_{9/2}^2$ (lower panel) at near minimum rotational frequency ($\hbar\omega$) for 
each configuration.}
\end{figure}

\begin{figure}
\includegraphics[angle=90,scale=.28,trim=0.0cm 0.0cm 4.0cm 0.0cm,clip=true]{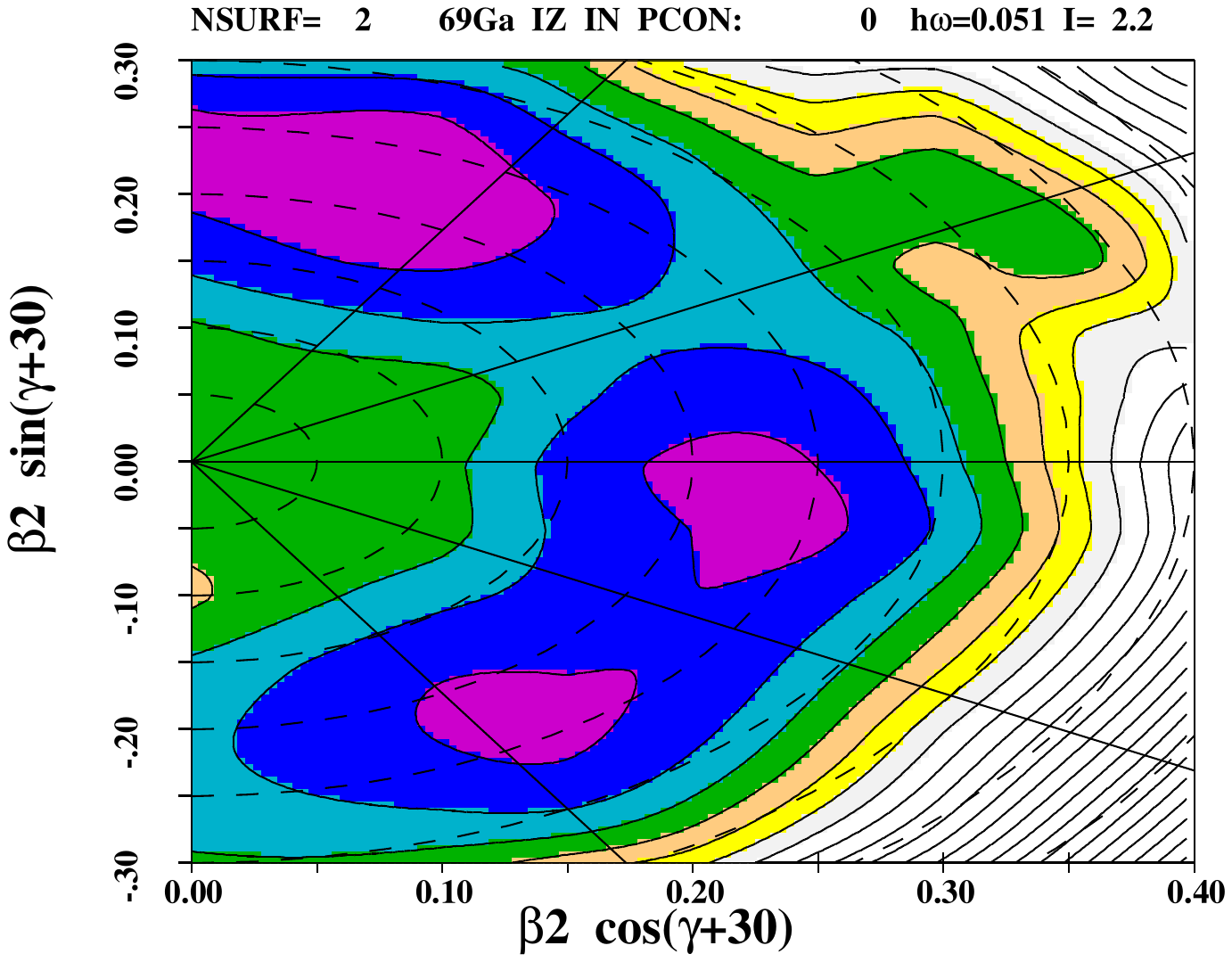}
\includegraphics[angle=0,scale=.25,trim=0.0cm 0.0cm 0.0cm 0.0cm,clip=true]{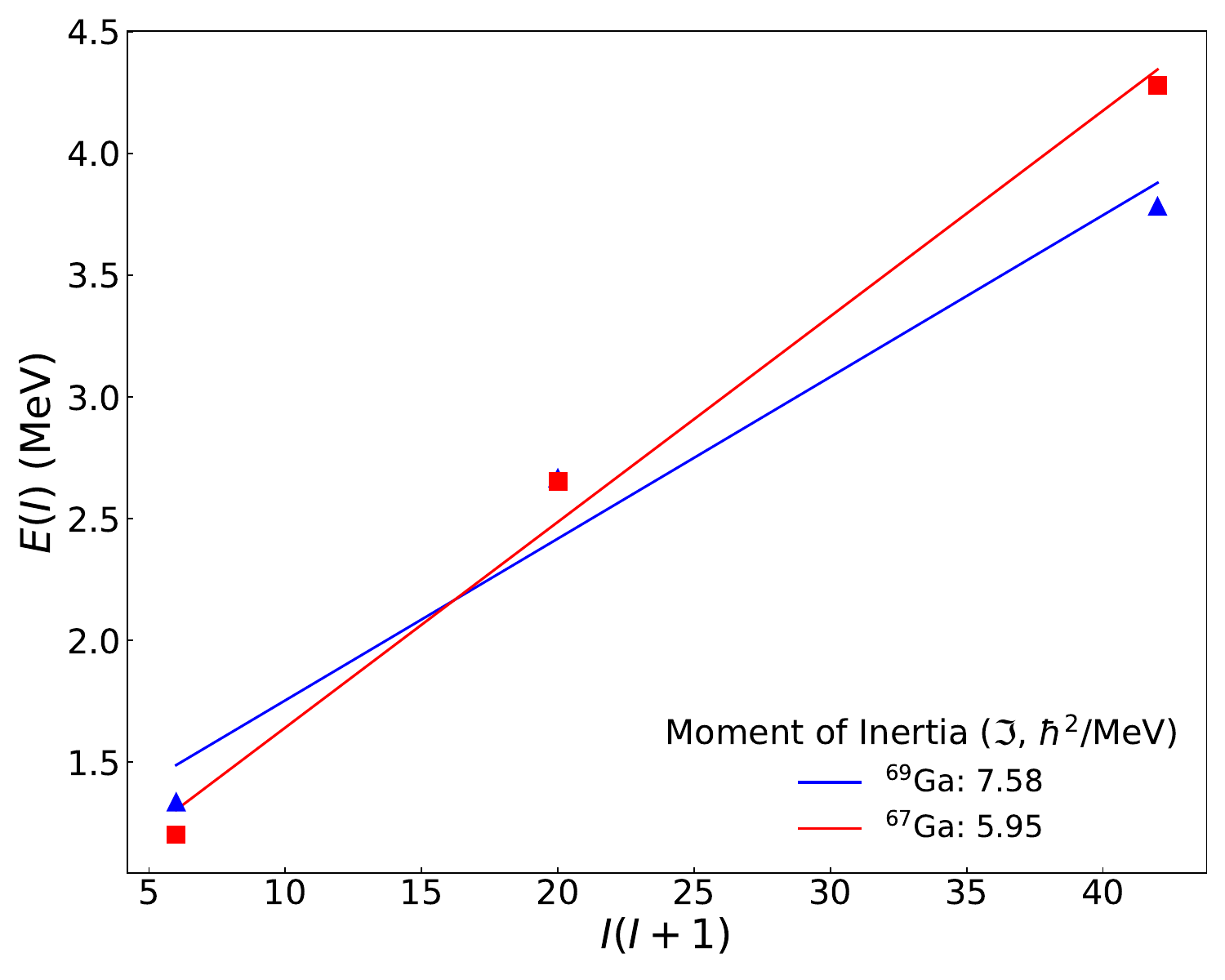}
\caption{\label{fig12} TRS plots for 3/2$^-$ ground-state band of the $^{69}$Ga nucleus corresponding to configurations $\pi p_{3/2}^1$ at near minimum rotational frequency ($\hbar\omega$) near the band-head (upper panel) 
and the corresponding $E(I)$ vs $I(I+1)$ plot along with that for $^{67}$Ga isotope (lower panel).}
\end{figure}

The $\pi f_{5/2}$ band, consisting of the yrast 5/2$^-$, 9/2$^-$, 13/2$^-$, 17/2$^-$, 21/2$^-$ and 25/2$^-$ states, is also a characteristic feature of 
the odd-$A$ (Ga) isotopes $^{67-71}$Ga \cite{Zho24,Zho21} and warrants a systematic comparison across the chain. The sequence in the present level scheme of $^{69}$Ga follows 
the reassignment of spin-parities of the 4077- and the 4526-keV states to 17/2$^-$ and 21/2$^-$ respectively. The characteristics of the $E(I)~vs~I(I+1)$ plot, 
the Moment of Inertia ($\Im$) therefrom, and the plot of the aligned angular momentum ($i_x$) against the rotational frequency ($\hbar\omega$) for the 
5/2$^-$ band, of $^{69}$Ga, fit well into the systematics of the similar structure observed in $^{67}$Ga and $^{71}$Ga, as illustrated in Fig.~10.
Similar to the 9/2$^+$ band, there is evidence of neutron-pair alignment herein and the corresponding frequency systematically evolves to lower values with increasing neutron number. 
Such compliance of the (5/2$^-$) band characteristics of $^{69}$Ga, to the existing trend, vindicates the (aforementioned) modification in the spin-parity assignments. 
The dominant particle configurations (Table~II) corresponding to the levels of this sequence in $^{69}$Ga, 
as emerging from the present shell model calculations, are commensurately dominated by an odd proton in the $f_{5/2}$ orbital while the neutron occupancy majorly 
consists of partitions with $\nu g_{9/2}^2$ for all levels of the sequence. The level energies are evidently better reproduced by the JUN45 interaction, albeit 
the deviations between the experimental and the theoretical values increase at higher excitations of the sequence that, presumably, follow the alignment. 
The TRS calculations for the deformation/ shape associated with this band are illustrated in Fig.~11 and exhibit considerable degree of $\gamma$-softness, at the 
band-head as well as around and after the alignment. This is characteristically distinct from the $g_{9/2}$ band and compulsively follows the respective 
particle configurations. \\

Similar character of $\gamma$-softness also underlies the 3/2$^-$ ground state band, as illustrated by the corresponding TRS plot in Fig.~12. The levels and transitions of this sequence
are from the previous studies \cite{Bak82} and validated in the present analysis. The MOI ($\Im$) value extracted from the slope of the $E(I)$ vs $I(I+1)$ 
plot, for this band, is close to that of the similar sequence in the neighboring $^{67}$Ga \cite{Zho24} isotope. 
The level energies (of this band) following shell model calculations using the JUN45 interaction agree superiorly (within $\lessapprox$ 100-keV) with the 
experimental values and the dominant configurations (Table~II) are based on $p_{3/2}^3$ as is expected from the systematics of the particle occupancies in these 
nuclei. This is substantially better than the overlap, between the experimental and the theoretical level energies, reported for the analogous sequence in the $^{71}$Ga \cite{Zho21} nucleus. 
However, a more stringent check on the validity of the particle configurations, and the wave functions they represent, would follow lifetime measurements 
of the levels and comparing the resulting transition probabilities with the calculated ones. \\

It may thus be construed that the shell model calculations are reasonably representative of the level structure associated with the negative parity bands in $^{69}$Ga 
based on $\pi p_{3/2}^1$ and $\pi f_{5/2}^1$, to the extent of the level energies and the associated particle configurations that overlap with the 
occupancies envisaged for the corresponding states. To the contrary, while the level energies of the $\pi g_{9/2}$ band are well reproduced in one or the other of the two (JUN45 and JJ44B) calculations, 
the particle configurations, for the lower states, largely fail to incorporate the odd-proton occupancy of the $\pi g_{9/2}$ orbital that steeply downslopes with ($\beta_2$) deformation
and competes with the $f_{5/2}$ (single particle energy) at $\beta \sim 0.3$. Interestingly, $\beta$ deformation of similar magnitude has been calculated for the 
band in the TRS framework. Further studies (outside the present scope) directed at level lifetime measurements, and the transition probabilities therefrom, are expected to facilitate better constraint
on the microscopic calculations for the evolving level structure of these transitional nuclei with considerable number of nucleons outside the closure.

\section{Conclusion}

The level structure of the $^{69}$Ga nucleus has been probed using heavy-ion induced reaction, to populate its excited states, and high resolution Compton suppressed 
HPGe clover array, as the detection system. The existing level scheme of the nucleus has been substantially extended based on the new $\gamma$-ray transitions along with 
their multipolarity assignments. A couple of existing assignments have been modified as well. Shell model calculations have been carried out for the
level energies and the overlap with their experimental values has been observed to vary depending on the choice of interaction. The level scheme is characterized by three 
band structures and their characteristics, vis-a-vis those of the similar structures in the neighboring isotopes, have been compared. These include their MOI and aligned angular momentum as well as
their shapes/ deformations that have been calculated in the TRS framework. The latter has exhibited largely prolate shape associated with the positive-parity $\pi g_{9/2}^1$ band and $\gamma$-soft potential minima
associated with the negative-parity bands based on $\pi f_{5/2}^1$ and $\pi p_{3/2}^1$ configurations. The different shapes of the nucleus could be ascribed to the occupancy of the respective orbitals 
underlying the sequences. The characteristics of deformation and collectivity in $^{69}$Ga are consistent with those of the isotopic chain and facilitate insights into their 
evolving behavior with changing neutron number in the valence space. One of such aspects is the evidence of octupole correlation that has been newly identified in $^{69}$Ga
and that is in agreement with the prediction of strong octupole effects at $N = 38$, based on the variation of energy of the 3$^-$ state in the neighboring even-even Zn isotopes.
This study is thus a comprehensive one, on the single particle and collective behavior of the $^{69}$Ga nucleus and brings out its transitional characteristics towards assuming those 
of the neutron rich isotopes. \\ 

\section*{Acknowledgments}

The authors wish to thank the staff associated with the
Pelletron Facility at IUAC, New Delhi, for their help and support during the experiment.
Help and support received from Mr. Kausik Basu (UGC-DAE CSR, Kolkata Centre), Dr. Anupriya Sharma (Himachal Pradesh University), Dr. Sutanu Bhattacharya (presently affiliated to The Hebrew University of Jerusalem), 
Dr. Sajad Ali (Government General Degree College at Pedong) and Dr. Subhendu Rajbanshi (Presidency University), during the experiment, are gratefully appreciated. 
The authors are indebted Dr. Rajamani Raghunathan for his support in the shell model calculations using the high performance computing facility, OJAS, at the Indore Centre of UGC-DAE CSR.
This work is partially supported by the Department of Science
and Technology, Government of India (No. IR/S2/PF-03/2003-II). SK acknowledges University Grants Commission, Government of India for fellowship under the NETJRF scheme. 
AD acknowledges Department of Science and Technology, Government of India for fellowship support under the INSPIRE scheme.

\bibliography{69Ga_rr}

\end {document}